\documentclass[fleqn,usenatbib]{mnras}

\usepackage{graphicx}
\usepackage{txfonts}
\usepackage[para,online,flushleft]{threeparttable}
\usepackage{xcolor}
\definecolor{redak}{rgb}{0.9,0.15,0.05}

\title[The cool supergiant problem and the HD limit]{The excess of cool supergiants from contemporary stellar evolution models defies the metallicity-independent Humphreys-Davidson limit}
\author[Gilkis, Shenar, Ramachandran, et al.]{Avishai~Gilkis$^{1}$\thanks{agilkis@tauex.tau.ac.il}, Tomer~Shenar$^{2}$\thanks{tomer.shenar@kuleuven.be}, Varsha~Ramachandran$^{3}$, Adam~S.~Jermyn$^{4}$, \and Laurent~Mahy$^{2}$, Lidia~M.~Oskinova$^{3}$, Iair~Arcavi$^{1,5}$ and Hugues~Sana$^{2}$
\\
$^{1}$ The School of Physics and Astronomy, Tel Aviv University, Tel Aviv 6997801, Israel\\
$^{2}$ Institute of Astrophysics, KU Leuven, Celestijnenlaan 200 D, 3001 Leuven, Belgium\\
$^{3}$ Institut f{\"u}r Physik und Astronomie, Universit{\"a}t Potsdam, Karl-Liebknecht-Str. 24 / 25, D-14476 Potsdam, Germany\\
$^{4}$ Center for Computational Astrophysics, Flatiron Institute, 162 5th Ave, New York, NY 10010, USA \\
$^{5}$ CIFAR Azrieli Global Scholars program, CIFAR, Toronto, Canada \\
}

\pubyear{2021}

\begin{document} 
\label{firstpage}
\pagerange{\pageref{firstpage}--\pageref{lastpage}}

\maketitle

\begin{abstract}
   The Humphreys-Davidson (HD) limit empirically defines a region of high luminosities ($ \log_{10} \left( L / \mathrm{L}_\odot \right) \gtrsim 5.5 $) and low effective temperatures ($ T_{\rm eff} \lesssim 20 \, {\rm kK} $) on the Hertzsprung-Russell Diagram in which hardly any supergiant stars are observed. Attempts to explain this limit through instabilities arising in near- or super-Eddington winds have been largely unsuccessful. Using modern stellar evolution we aim to re-examine the HD limit, investigating the impact of enhanced mixing on massive stars. We construct grids of stellar evolution models appropriate for the Small and Large Magellanic Clouds (SMC, LMC), as well as for the Galaxy, spanning various initial rotation rates and convective overshooting parameters. Significantly enhanced mixing apparently steers stellar evolution tracks away from the region of the HD limit. To quantify the excess of over-luminous stars in stellar evolution simulations we generate synthetic populations of massive stars, and make detailed comparisons with catalogues of cool ($T_\mathrm{eff} \le 12.5\,\mathrm{kK}$) and luminous ($ \log_{10} \left( L / \mathrm{L}_\odot\right) \ge 4.7 $) stars in the SMC and LMC. We find that adjustments to the mixing parameters can lead to agreement between the observed and simulated red supergiant populations, but for hotter supergiants the simulations always over-predict the number of very luminous ($ \log_{10} \left( L / \mathrm{L}_\odot\right) \ge 5.4 $) stars compared to observations. The excess of luminous supergiants decreases for enhanced mixing, possibly hinting at an important role mixing has in explaining the HD limit. Still, the HD limit remains unexplained for hotter supergiants.
\end{abstract}


%

\section{Introduction}
\label{sec:intro}

\begin{figure*}
\centering
\begin{tabular}{cc}
\includegraphics[width=0.475\textwidth]{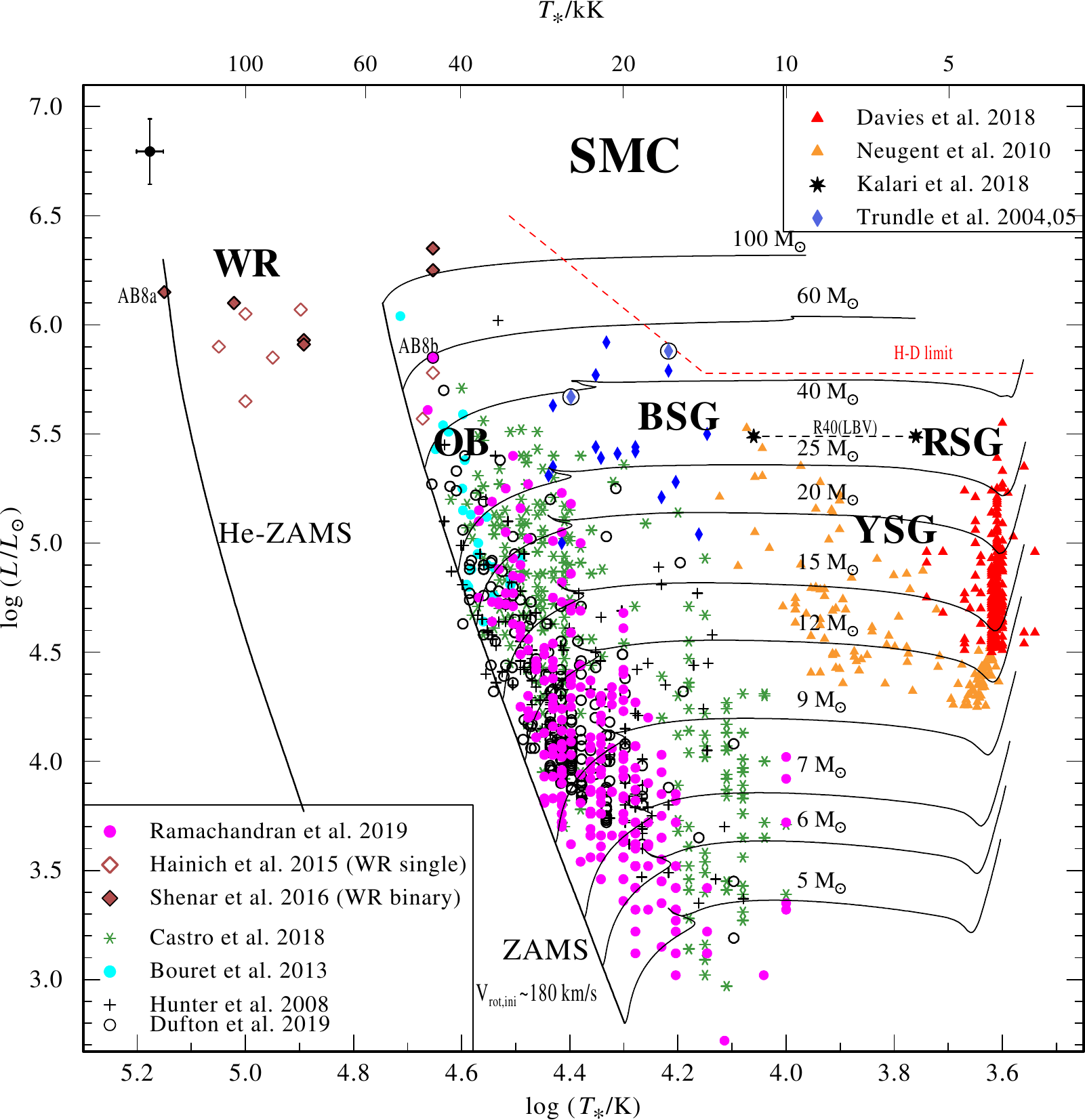} & 
\includegraphics[width=0.475\textwidth]{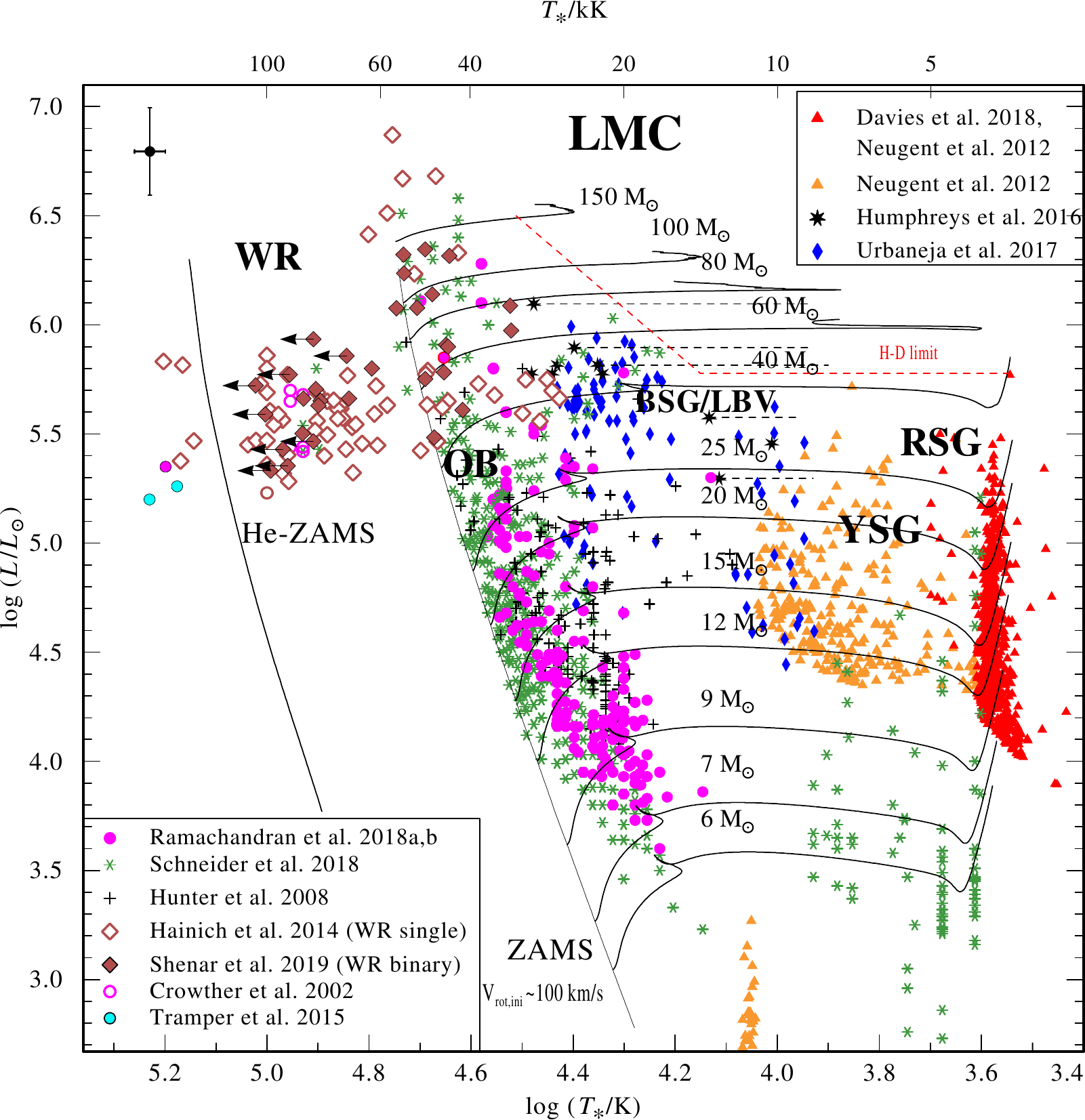} 
\end{tabular}
\caption{HRD positions of populations of massive
stars in the SMC (left panel) and LMC (right panel), based on analyses of apparently single and binary WR stars \citep{Hainich2014,Hainich2015,Shenar2016,Shenar2019,Crowther2002,Tramper2015}, YSGs \citep{Neugent2010,Neugent2012}, RSGs \citep{HDrevisited,Neugent2012}, LBVs \citep{humphreys_social_2016,kalari_how_2018}, BSGs \citep{trundle_understanding_2004,trundle_understanding_2005,urbaneja_lmc_2017}, and populations of OB-type stars \citep{ramachandran_stellar_2018a,ramachandran_stellar_2018b,ramachandran_testing_2019,schneider_vltflames_2018,hunter_vlt_2008,castro_spectroscopic_2018,dufton_census_2019,bouret_massive_2013}. The typical error bar is shown in the top left corner. Evolutionary tracks (black solid lines), accounting for rotation with $\varv_{\mathrm{rot, init}} \,\approx$\,100 km\,s$^{-1} $ for the LMC and $\approx$\,180 km\,s$^{-1} $ for the SMC \citep{brott_rotating_2011,kohler_evolution_2015}, are labeled with their initial mass. The red dashed lines mark the observational limit reported by \citet{HD1979}. The diagrams are largely complete for supergiants with $T_{\rm eff} \lesssim 12.5\,\mathrm{kK}$, but not necessarily complete for hotter stars.}
\label{fig:Varsha}
\end{figure*}

The upper-right part of the Hertzsprung-Russell diagram (HRD) features a stark absence of observed stars (Fig.\,\ref{fig:Varsha}), a phenomenon termed the Humphreys-Davidson (HD) limit \citep{HD1979}. While the luminous blue variables (LBVs) venture into this region during outbursts, with a few exceptions, cool supergiants (CSGs), which comprise red, yellow, and blue supergiants (RSGs, YSGs, BSGs)  with effective temperatures $T_{\rm eff} \lesssim 12.5\,\mathrm{kK}$ and luminosities $\log_{10} \left(L_{\rm max} / \mathrm{L}_\odot\right) \gtrsim 5.5$ (i.e., initial masses $M_{\rm i} \gtrsim 30\,\mathrm{M}_\odot$) are not observed. In contrast, hundreds of main sequence progenitors with $M_{\rm i} \gtrsim 30\,\mathrm{M}_\odot$, ranging all the way to $\approx 150\,\mathrm{M}_\odot$ and perhaps more, were directly observed \citep{Crowther2010, Bestenlehner2011, Almeida2017, Shenar2017, Tehrani2019, Mahy2020}. This implies two possibilities: either stars with $M_{\rm i} \gtrsim 30\,\mathrm{M}_\odot$ skip the CSG phase altogether, or they experience this phase very briefly, making it observationally rare. 

Despite various attempts, no evolutionary models are currently capable of reproducing the observed absence of stars beyond the HD limit \citep*{HDrevisited, Schootemeijer2018}. The HD limit has consequences not only for our understanding of the evolution of the progenitors of Wolf-Rayet (WR) stars and black holes (BHs), but also for estimates of the likelihood of binary interaction in the upper-mass end of stars. The reliability of our predictions of gravitational-wave (GW) events are thus severely limited as long as the HD limit has not been sufficiently understood.

It is commonly assumed that the HD limit is a result of radiative instability, tracing a region in which stars reach their Eddington luminosity and become LBVs \citep[e.g.,][]{LamersFitzpatrick1988, GlatzelKiriakidis1993}. However, attempts to prove this by means of direct radiative transfer calculations in stellar models have had only limited success.  \citet{Ulmer1998} showed that the so-called modified Eddington limit mimics the shape of the HD limit, but that it lies roughly one magnitude above the observed HD limit. More recently, \citet{Sanyal2017} showed that stars with $M_{\rm i} \gtrsim30\,\mathrm{M}_\odot$ and a solar metallicity content ($Z = Z_\odot$) reach the Eddington limit in their interiors and undergo envelope inflation. However, both these studies and others clearly predict that $L_{\rm max}$ should grow with decreasing $Z$. This is because the opacities in the interior of stars are strongly correlated with the content of metals within them. The lower $Z$ is, the weaker the outward radiative pressure becomes. Hence, from this perspective, environments with a lower ambient $Z$ are expected to allow the stability of more massive stars, and higher $L_{\rm max}$.

This prediction does not seem to be confirmed by observations. \citet{HD1979} originally reported a maximum RSG luminosity at around $\log_{10}\left( L / \mathrm{L}_\odot\right) \approx 5.7$, later revised slightly downwards to $5.6$  by \citet{Levesque2005}. Similar values are reported by \citet{Massey2016} and \citet*{Drout2012} for the super-solar metallicity environment of the Andromeda galaxy (M31) and the approximately solar $Z$-environment of the Triangulum galaxy (M33). 

This trend continues with the Small and Large Magellanic Clouds (SMC, LMC), which have metallicities of $Z \approx 0.15, 0.4\,Z_\odot$, respectively \citep{Korn2000, Hunter2007, Trundle2007}. Figure~\ref{fig:Varsha} shows the HRD positions of analysed massive stars in the SMC and LMC, adapted from \citet{ramachandran_stellar_2018b, ramachandran_testing_2019}. These populations are thought to be complete for the WRs as well as the luminous ($ \log_{10} \left( L / \mathrm{L}_\odot\right) \gtrsim 4.7 $) YSGs and RSGs, but far from complete for the OB-type stars. Figure~\ref{fig:Varsha} illustrates strongly the absence of RSGs with $ \log_{10} \left( L / \mathrm{L}_\odot\right) \gtrsim 5.5 $ in both galaxies, with the exception of two peculiar objects in the LMC: {HD~33579} \citep{Wolf1972} and {WOH~G~64} \citep{Ohnaka2008}.

\citet{HDrevisited} performed a statistical comparison of analyzed RSG populations in the SMC and LMC, and showed compelling evidence that $ \log_{10} \left( L_{\rm max} / \mathrm{L}_\odot\right) \lesssim 5.5 $ in both galaxies. Moreover, no evidence is found for higher $L_{\rm max}$ in the SMC compared to the LMC. \citet{HDrevisited} illustrated why this is very likely a genuine physical fact rather than a statistical one, and how this stands in tension with recent rotating and non-rotating Geneva evolution models \citep{Ekstrometal2012,Georgyetal2013}. 

The challenge of explaining the HD limit solely by radiative instabilities might imply that other processes are involved. It was already recognized in the past that rotationally induced mixing can strongly hinder the redward evolution of massive stars towards the RSG phase \citep[e.g.,][]{Maeder2000}. However, initial rotation rates in the excess of $\approx 200\,\mathrm{km}\,\mathrm{s}^{-1}$ (at which strong rotational mixing starts) seem to be inconsistent with observed rotation rates of massive stars \citep{RamirezAgudelo2015,RamirezAgudelo2017,ramachandran_testing_2019}, and models using realistic rotation rates do not reproduce the apparently $Z$-independent HD limit \citep{HDrevisited}.

\citet{HigginsVink2020} recently explored this problem by investigating the impact of mixing parameters on RSG stellar models. They found that enhanced semi-convection can reproduce the observed HD limit. However, a solution to the HD limit problem needs to consider YSGs and cool BSGs as well, as we argue in our study. 

In this paper we tackle the puzzle of the HD limit from a different angle, quantifying the duration spent beyond the HD limit and the expected number of stars for various stellar evolution models. For this purpose we study models with enhanced mixing and construct synthetic populations based on these models. Moreover, unlike previous studies that considered RSGs alone, we consider simultaneously RSGs, YSGs, and cool BSGs ($T_\mathrm{eff} \lesssim 12.5\,\mathrm{kK}$). As our study shows, considering CSGs as a whole is vital for correctly assessing the discrepancy between models and observations.

The paper is organized as follows. In Section \ref{sec:method} we describe our numerical approach. In Section \ref{sec:results} we show the effect of enhanced mixing on stellar evolution tracks. In Section \ref{sec:popsynvsobs} we make comparisons between observations and synthetic populations that we generate from our stellar evolution tracks. In Section \ref{sec:assumptions} we examine the impact of modelling assumptions. We discuss our results in relation to previous studies, and summarise, in Section \ref{sec:discussion}.


\section{Numerical method}
\label{sec:method}

We use the Modules for Experiments in Stellar Astrophysics code (\texttt{MESA}, version 10398, \citealt{Paxton2011,Paxton2013,Paxton2015,Paxton2018}) to evolve stellar models with $39$ different zero-age main sequence (ZAMS) masses between $M_\mathrm{ZAMS}=4\,\mathrm{M}_\odot$ and $M_\mathrm{ZAMS}=107\,\mathrm{M}_\odot$. Three different initial compositions are used, appropriate for SMC, LMC and Milky Way (MW) stars, as listed in Table \ref{Table:Z} (following \citealt{Hainich2019}).
\begin{table}
\centering
\caption{Initial compositions in terms of mass fractions for stellar evolution calculations.}
\begin{threeparttable}
\begin{tabular}{c|ccc}
\hline
\hline
& SMC & LMC & MW \\
\hline
H & $0.75328$ & $0.74392$ & $0.72031$ \\
He & $0.24448$ & $0.25072$ & $0.26646$ \\
C & $0.00039621$ & $0.00094807$ & $0.0023401$ \\
N & $0.00011606$ & $0.00027772$ & $0.00068549$ \\
O & $0.00096042$ & $0.0022981$ & $0.0056725$ \\
Ne & $0.00021246$ & $0.00050838$ & $0.0012548$ \\
Mg & $0.00011456$ & $0.00027412$ & $0.00067661$ \\
Si & $0.00012179$ & $0.00029144$ & $0.00071934$ \\
S & $0.000056647$ & $0.00013555$ & $0.00033457$ \\
Ar & $0.000013447$ & $0.000032178$ & $0.000079424$ \\
Ca & $0.00001175$ & $0.000028117$ & $0.0000694$ \\
Fe & $0.00023666$ & $0.0005663$ & $0.0013978$ \\
\hline
$Z$ & $0.00224$ & $0.00536$ & $0.01323$ \\
\hline
\end{tabular}
\footnotesize
\end{threeparttable}
\label{Table:Z}
\end{table}

\subsection{Microphysics}
\label{subsec:micphys}

The equation of state employed by \texttt{MESA} is a blend of the following equations of state: OPAL \citep{Rogers2002}, SCVH \citep*{Saumon1995}, HELM \citep{Timmes2000}, and PC \citep{Potekhin2010}. Radiative opacities are primarily from OPAL \citep{Iglesias1993, Iglesias1996}, with low-temperature data from \citet{Ferguson2005} and the high-temperature, Compton-scattering dominated regime according to \citet{Buchler1976}. Electron conduction opacities follow \citet{Cassisi2007}.

We  use  the  built-in \texttt{MESA} nuclear reaction network \texttt{approx21}. The Joint Institute for Nuclear Astrophysics REACLIB reaction rates \citep{Cyburt2010} are used, with additional tabulated weak reaction rates (\citealt*{Fuller1985}; \citealt{Oda1994, Langanke2000}) and screening via the prescriptions of \citet{Salpeter1954}, \citet*{Dewitt1973}, \citet{Alastuey1978} and \citet{Itoh1979}. The formulae of \citet{Itoh1996} are used for thermal neutrino loss rates.

\subsection{Wind mass loss}
\label{subsec:winds}

For cool phases ($T_\mathrm{eff} \le 10\,000\,\mathrm{K}$), the mass-loss prescription of \cite*{deJager1988} is employed, regardless of the surface hydrogen mass fraction $X_\mathrm{s}$. Wind mass loss is according to \cite*{Vink2001} for hot ($T_\mathrm{eff} \ge 11\,000\,\mathrm{K}$) hydrogen-rich phases ($X_\mathrm{s}\ge 0.4$) of the evolution, most notably the main sequence. If the surface hydrogen mass fraction is low but non-negligible ($0.1 \le X_\mathrm{s} < 0.4$) then the empirical mass-loss rate relation of \cite{NL00} is used. For hot hydrogen-deficient models ($X_\mathrm{s} < 0.1$) we follow \cite{Yoon2017} and \cite{Woosley2019}, whose prescriptions are based on the mass-loss rates of \cite{Hainich2014} and \cite*{TSK2016}. For $10\,000\,\mathrm{K} < T_\mathrm{eff} < 11\,000\,\mathrm{K}$ we interpolate between the two regimes described above. The mass-loss rate is multiplied by a factor $\eta_\mathrm{w}$. We mostly use $\eta_\mathrm{w}=1$, but in one set of models we explore the impact of boosted mass loss on our results by taking $\eta_\mathrm{w}=2$ (Section\,\ref{subsec:massloss}) throughout the entire evolution. We note that the metallicity dependence of the mass-loss rate during the main sequence, which follows \cite{Vink2001}, is $\dot{M}\propto Z^{0.85}$. The fit by \cite{deJager1988}, which prevails during most of the CSG phases, has no implicit dependence on metallicity.

\subsection{Mixing and rotation}
\label{subsec:mixrot}

The models have initial rotation velocities at the equator of $V_\mathrm{i} = 100$, $200$, and $300\,\mathrm{km}\,\mathrm{s}^{-1}$. The shellular approximation where the angular velocity is constant over isobars \citep{MM1997} is used for rotating models in \texttt{MESA} \citep{Paxton2013}. Mixing processes induced by rotation are implemented in a diffusion approximation \citep{Paxton2013} and principally follow \cite*{Hegeretal2000}. Transport of angular momentum and chemical mixing caused by internal magnetic fields follows the new prescription of \cite*{TSF}.

The efficiency of rotational mixing is set by two parameters, the ratio of the turbulent viscosity to the diffusion coefficient $f_c$ (input parameter \texttt{am\_D\_mix\_factor} in \texttt{MESA}), and the ratio between the actual molecular weight gradient and the value used for computing the mixing coefficients $f_\mu$ (input parameter \texttt{am\_gradmu\_factor} in \texttt{MESA}). These parameters are calibrated to give the observed nitrogen enhancement for evolved stars. We use the calibration of \cite{Hegeretal2000}, $f_c = 1/30$ and $f_\mu=0.05$ for most models. However, as discussed by \cite{ChieffiLimongi2013}, this calibration is not unique (also \citealt*{PTE2012}). In one set of models we use alternative values of $f_c = 0.2$ and $f_\mu=1$ \citep{ChieffiLimongi2013}.

Convective regions are defined by the Ledoux criterion (except for one set of models which employs the Schwarzschild criterion) and treated according to \cite*{MLT} with a mixing-length parameter of $\alpha_\mathrm{MLT}=1.5$. Semiconvective mixing in regions which are Ledoux stable but Schwarzschild unstable follows \cite*{Langer1983} with an efficiency parameter $\alpha_\mathrm{sc}=1$ or $\alpha_\mathrm{sc}=100$. Thermohaline mixing is according to \cite*{thermohaline} with an efficiency parameter of $\alpha_\mathrm{th}=1$. We use the so-called MLT++ implementation for efficient energy transport in convective regions \citep{Paxton2013}.

Mixing above the convective core boundary is extended in two approaches. First, for most models, a step overshoot approach is taken (e.g. \citealt{ShavivSalpeter1973,MM1987}), where the convective region is extended by a fraction $\alpha_\mathrm{ov}$ of the pressure scale height $H_P$. The lowest value for $\alpha_\mathrm{ov}$ that we use is $\alpha_\mathrm{ov}=0.1$ (e.g. \citealt{Ekstrometal2012,HigginsVink2020}), followed by $\alpha_\mathrm{ov}=0.335$ as calibrated by \cite{brott_rotating_2011}, with higher values starting at $\alpha_\mathrm{ov}=0.5$ (e.g. \citealt{Vinketal2010,HigginsVink2019}), and additional higher values of $\alpha_\mathrm{ov}=0.8$, $\alpha_\mathrm{ov}=1$ and $\alpha_\mathrm{ov}=1.2$\footnote{These parameter values might not be physical, but rather used as proxy to investigate the issue at hand.}. In a second approach we use an exponential core overshooting (e.g. \citealt{Herwig2000}), where the mixing efficiency decays smoothly outside the core, rather than dropping abruptly as in the step overshoot approach. We follow the prescription of \cite*{JTC} to compute the fraction $f_\mathrm{ov}$ used for the decay scale,
\begin{eqnarray}
f_\mathrm{ov} = \frac{2}{\ln 150 - 2 \ln \left(v_\mathrm{c} / c_\mathrm{s}\right) + \left(\frac{5}{2} \left(H_P/r_\mathrm{c}\right) - 1\right)},
\label{eq:fov}
\end{eqnarray}
where $v_\mathrm{c} / c_\mathrm{s}$ is the mass-averaged convective core Mach number, $H_P$ is taken at the top of the core, and $r_\mathrm{c}$ is the convective core radius. The value of $f_\mathrm{ov}$ is updated according to equation (\ref{eq:fov}) at the end of every evolution step. This prescription corresponds to a meridional circulation driven by anisotropy in the heat flux emerging from the convective core. The anisotropy is rotationally-induced, but because the convective turnover time is so long in the core this effect saturates at slower angular velocities than any we consider here.


\section{Stellar evolution tracks}
\label{sec:results}

A total of $5967$ models were evolved, for $39$ initial masses ($M_\mathrm{ZAMS} = 4$, $5$, $6$, $7$, $8$, $9$, $10$, $11$, $12$, $13$, $14$, $15$, $16$, $17$, $19$, $20$, $22$, $23$, $25$, $27$, $29$, $31$, $33$, $36$, $39$, $42$, $45$, $48$, $52$, $56$, $60$, $64$, $69$, $74$, $80$, $86$, $92$, $99$, and $107\,\mathrm{M}_\odot$), $3$ initial rotation velocities ($V_\mathrm{i} = 100$, $200$, and $300\,\mathrm{km}\,\mathrm{s}^{-1}$), $3$ compositions (SMC, LMC, and MW), and $17$ sets of modelling assumptions as detailed in Table \ref{tab:modelassumptions}. The evolution reached the end of core helium burning for the lower masses ($M_\mathrm{ZAMS} < 10\,\mathrm{M}_\odot$) or core carbon burning for the higher masses ($M_\mathrm{ZAMS} \ge 10\,\mathrm{M}_\odot$). We present several sub-sets of the results to highlight our main findings.
\begin{table}
\centering
\caption{Modelling assumptions for all sets of models computed.}
\begin{threeparttable}
\begin{tabular}{ccccccc}
\hline
\hline
\# & stability & overshooting & $\alpha_\mathrm{sc}$ & $f_c$ & $f_\mu$ & $\eta_\mathrm{w}$\\
& criterion & & & & &\\
\hline
$1$ & Ledoux & $\alpha_\mathrm{ov}=0.1$ & $1$ & $1/30$ & $0.05$ & $1$ \\
$2$ & Ledoux & $\alpha_\mathrm{ov}=0.335$ & $1$ & $1/30$ & $0.05$ & $1$ \\
$3$ & Ledoux & $\alpha_\mathrm{ov}=0.5$ & $1$ & $1/30$ & $0.05$ & $1$ \\
$4$ & Ledoux & $\alpha_\mathrm{ov}=0.8$ & $1$ & $1/30$ & $0.05$ & $1$ \\
$5$ & Ledoux & $\alpha_\mathrm{ov}=1.0$ & $1$ & $1/30$ & $0.05$ & $1$ \\
$6$ & Ledoux & $\alpha_\mathrm{ov}=1.2$ & $1$ & $1/30$ & $0.05$ & $1$ \\
$7$ & Ledoux & $f_\mathrm{ov,JTC}$ & $1$ & $1/30$ & $0.05$ & $1$ \\
\hline
$8$ & Ledoux & $\alpha_\mathrm{ov}=0.1$ & $100$ & $1/30$ & $0.05$ & $1$ \\
$9$ & Ledoux & $\alpha_\mathrm{ov}=0.335$ & $100$ & $1/30$ & $0.05$ & $1$ \\
$10$ & Ledoux & $\alpha_\mathrm{ov}=0.5$ & $100$ & $1/30$ & $0.05$ & $1$ \\
$11$ & Ledoux & $\alpha_\mathrm{ov}=0.8$ & $100$ & $1/30$ & $0.05$ & $1$ \\
$12$ & Ledoux & $\alpha_\mathrm{ov}=1.0$ & $100$ & $1/30$ & $0.05$ & $1$ \\
$13$ & Ledoux & $\alpha_\mathrm{ov}=1.2$ & $100$ & $1/30$ & $0.05$ & $1$ \\
$14$ & Ledoux & $f_\mathrm{ov,JTC}$ & $100$ & $1/30$ & $0.05$ & $1$ \\
\hline
$15$ & Schwarzschild & $\alpha_\mathrm{ov}=0.335$ & $-$ & $1/30$ & $0.05$ & $1$ \\
$16$ & Ledoux & $\alpha_\mathrm{ov}=0.335$ & $100$ & $0.2$ & $1.0$ & $1$ \\
$17$ & Ledoux & $\alpha_\mathrm{ov}=0.335$ & $100$ & $1/30$ & $0.05$ & $2$ \\
\end{tabular}
\footnotesize
\end{threeparttable}
\label{tab:modelassumptions}
\end{table}
\begin{figure*}
\centering
\begin{tabular}{cc}
\includegraphics[width=0.475\textwidth]{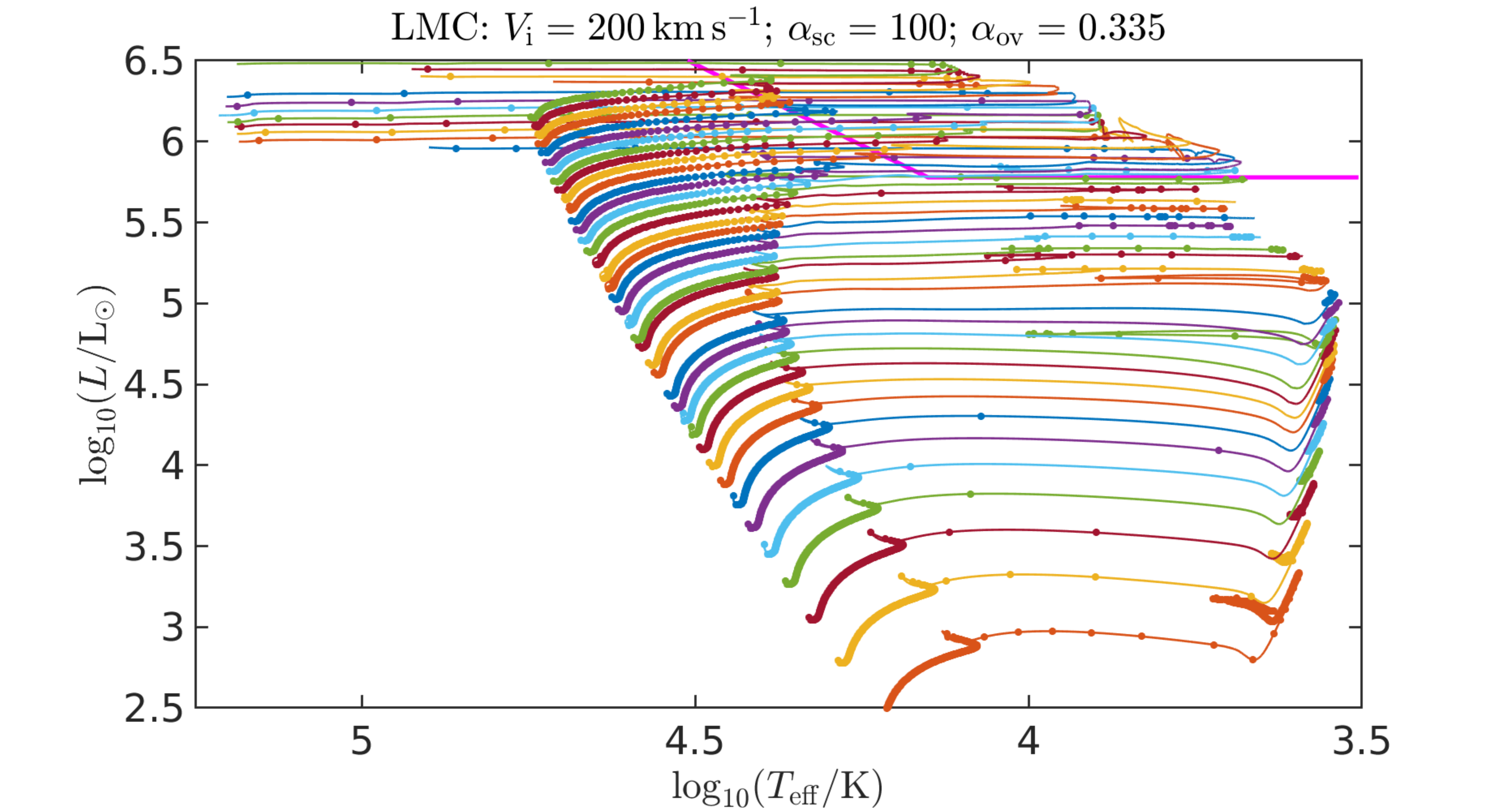} & 
\includegraphics[width=0.475\textwidth]{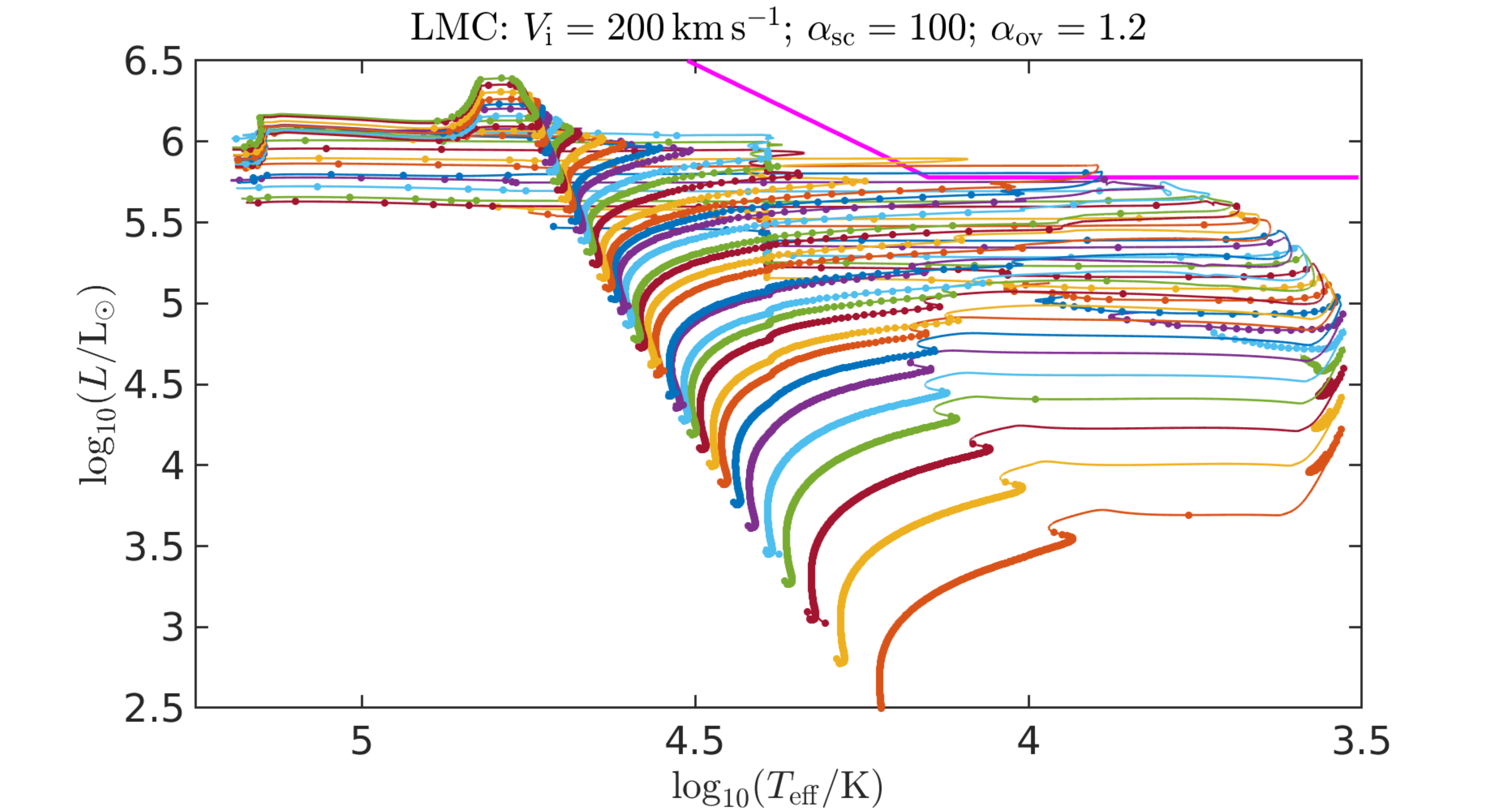} \\ 
\includegraphics[width=0.475\textwidth]{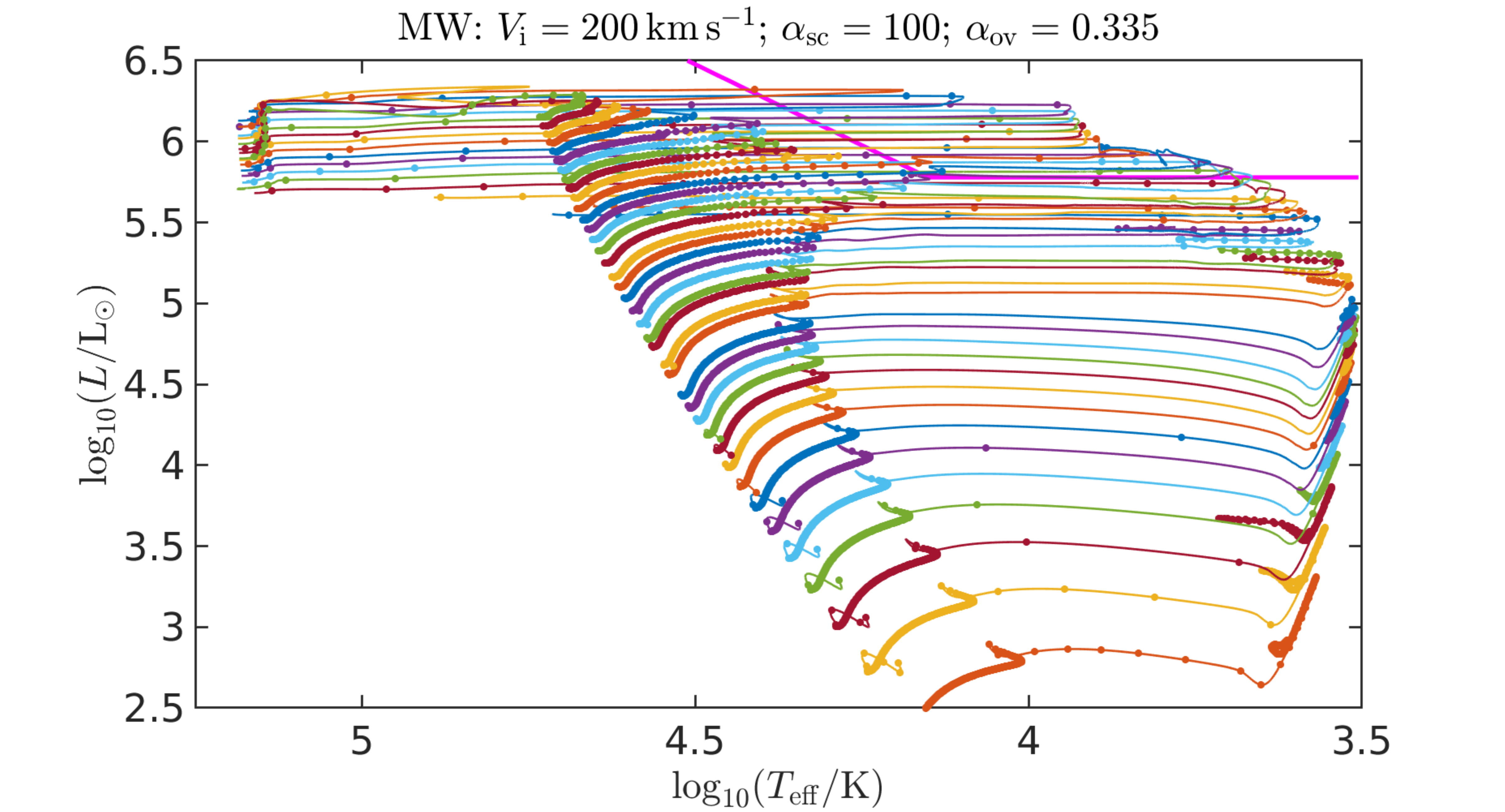} & 
\includegraphics[width=0.475\textwidth]{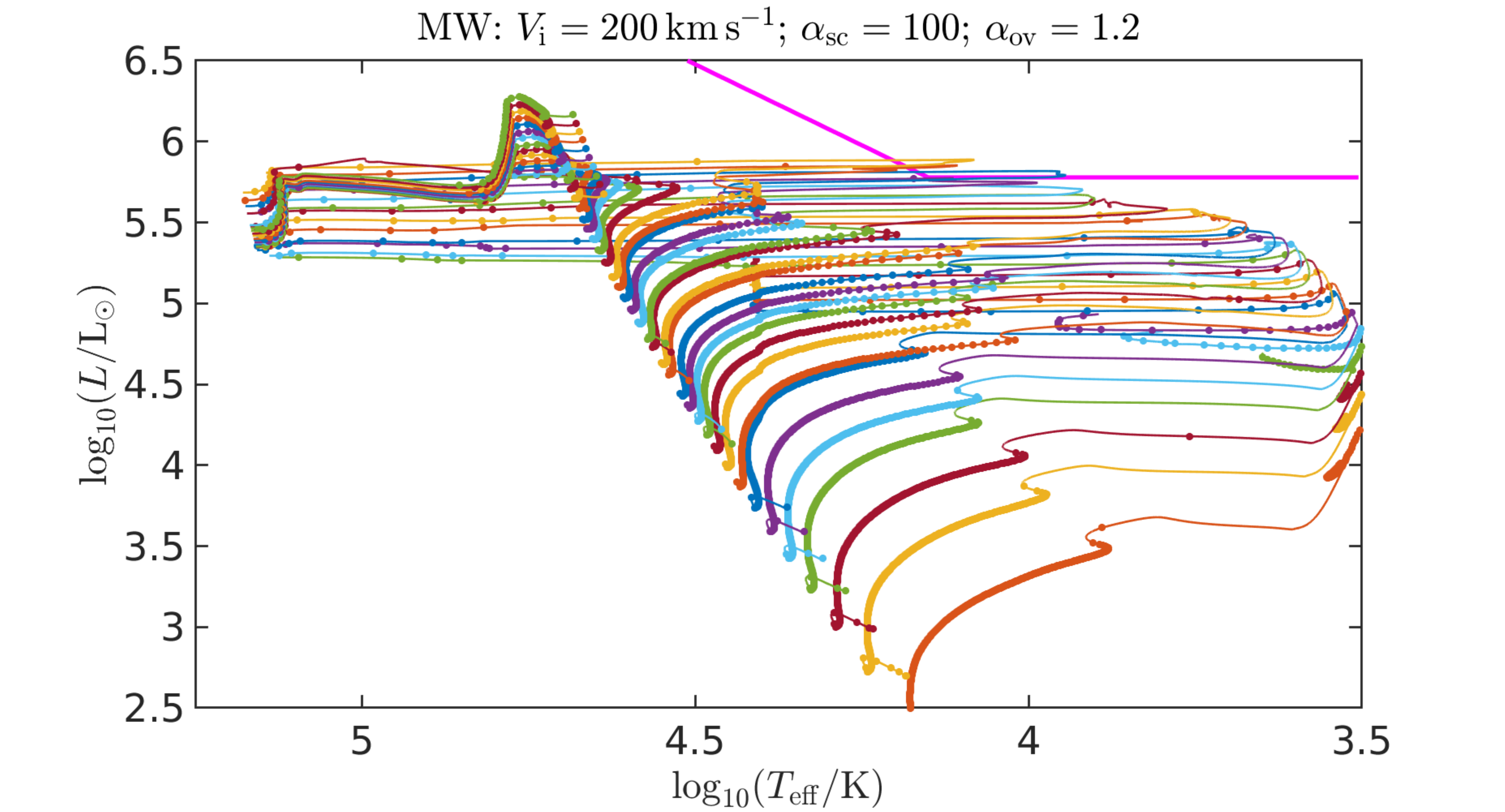} \\ 
\end{tabular}
\caption{Hertzsprung-Russell diagrams for models with a semiconvective mixing efficiency of $\alpha_\mathrm{sc}=100$ and an overshoot parameter of $\alpha_\mathrm{ov} = 0.335$ (left) and $\alpha_\mathrm{ov} = 1.2$ (right) for LMC (top) and MW (bottom) initial compositions, with an initial rotation velocity of $200\,\mathrm{km}\,\mathrm{s}^{-1}$. Initial masses are in the range $4\,\mathrm{M}_\odot \le M_\mathrm{ZAMS} \le 107\,\mathrm{M}_\odot$. Models every $50\, 000 \, \mathrm{yr}$ are marked. The thick magenta line maps the HD limit.}
\label{fig:HR1}
\end{figure*}
\begin{figure*}
\centering
\begin{tabular}{cc}
\includegraphics[width=0.475\textwidth]{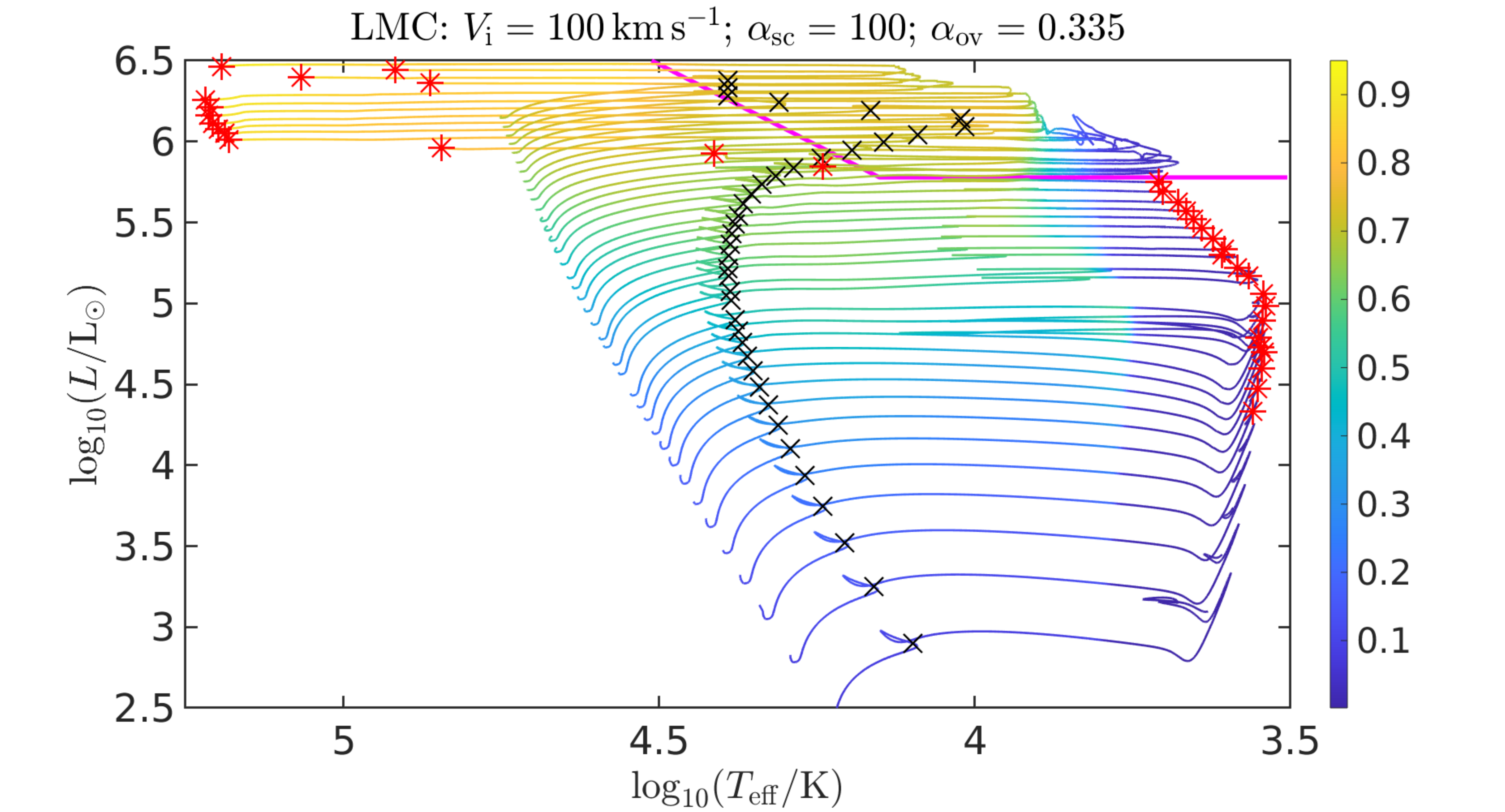} & 
\includegraphics[width=0.475\textwidth]{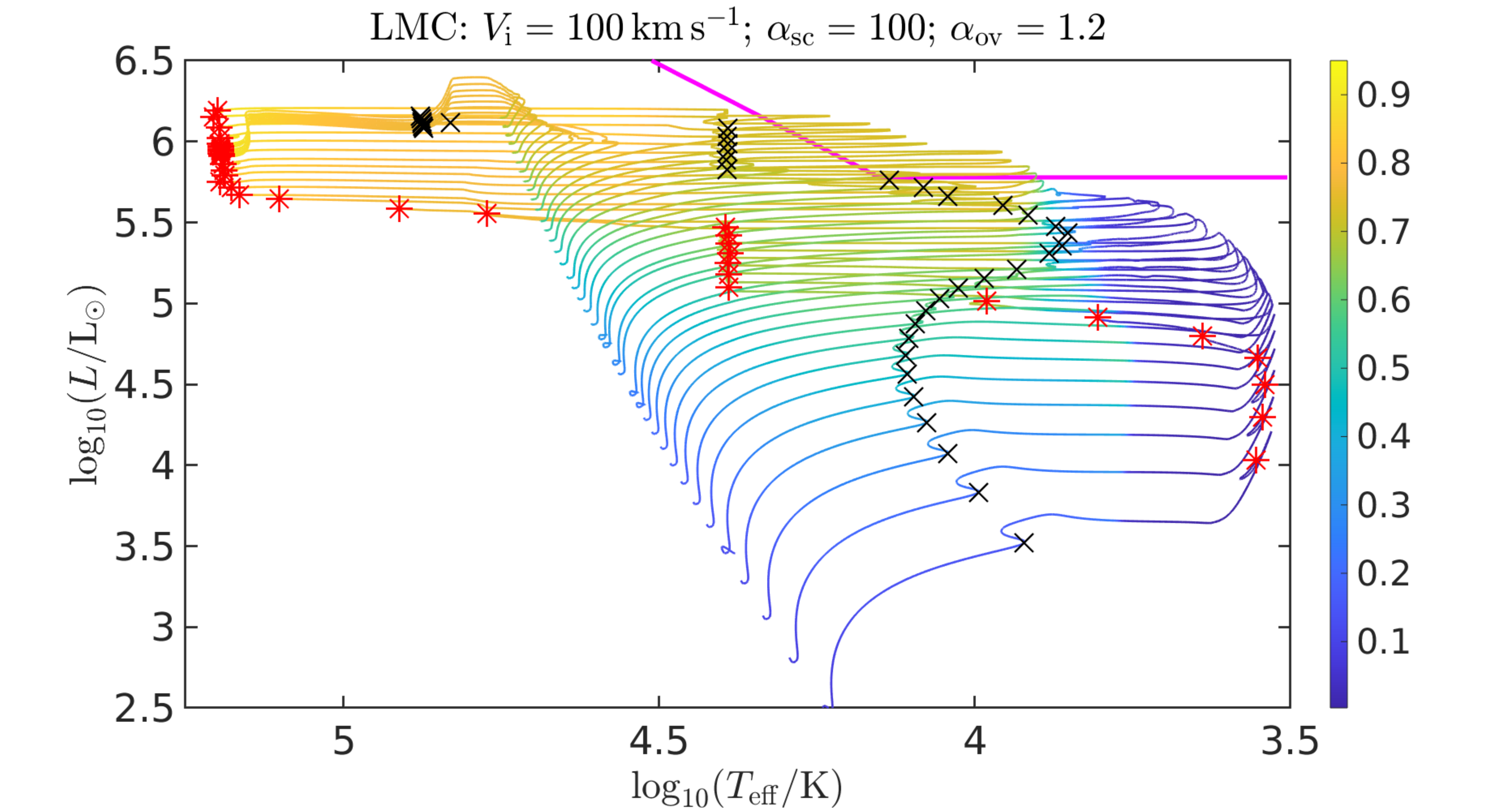} \\ 
\includegraphics[width=0.475\textwidth]{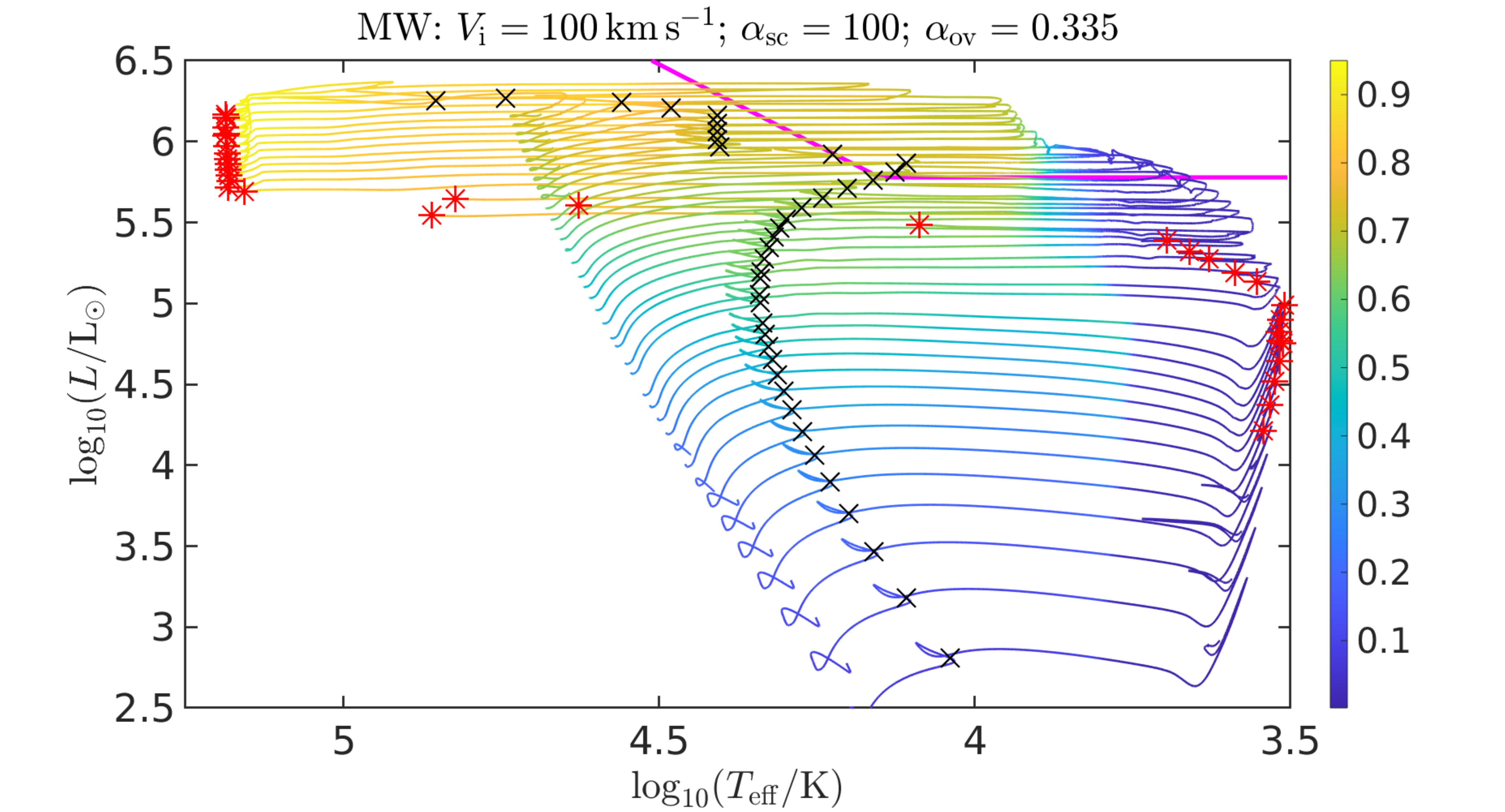} & 
\includegraphics[width=0.475\textwidth]{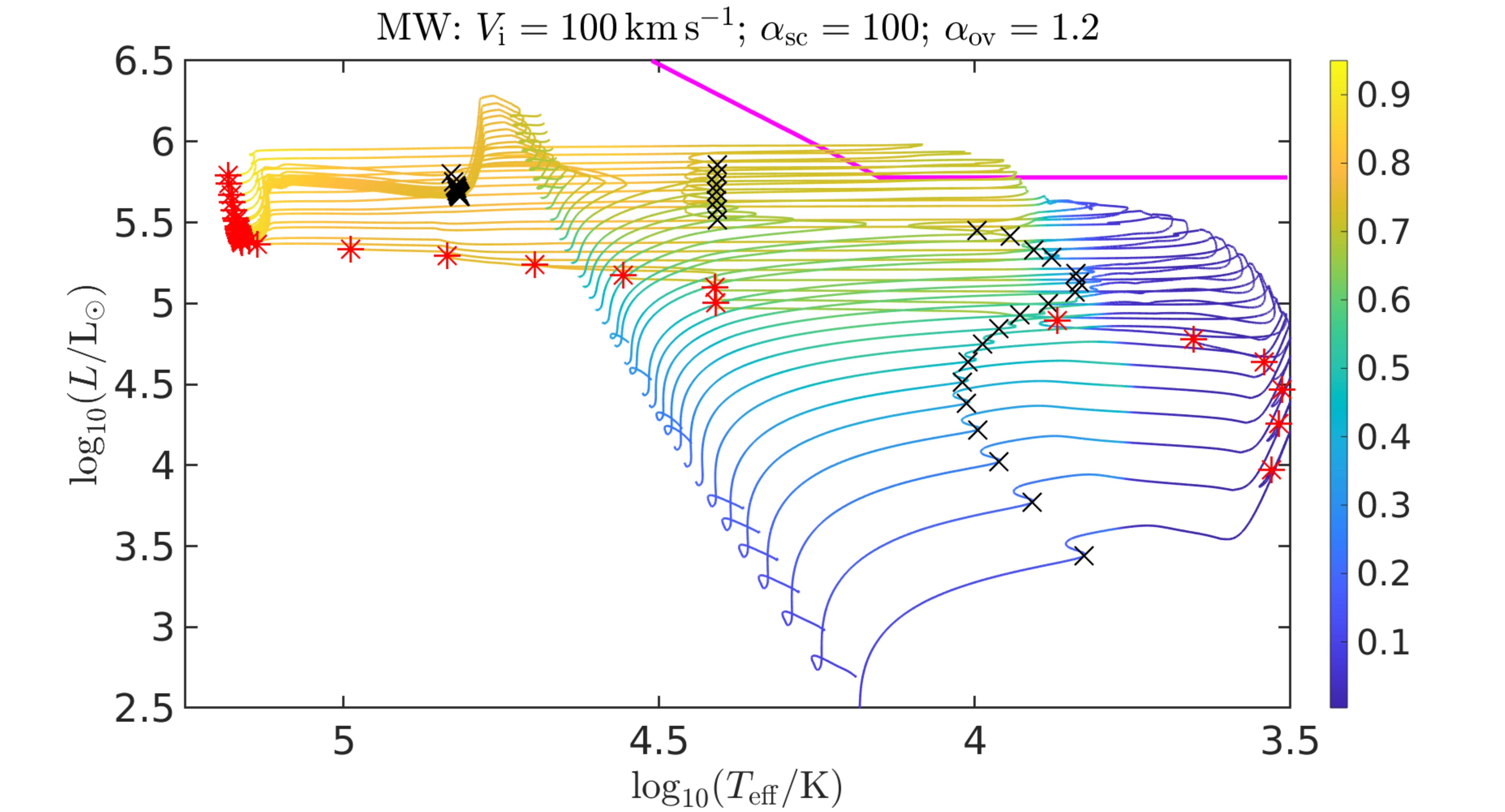} \\ 
\end{tabular}
\caption{ Hertzsprung-Russell diagrams for models with a semiconvective mixing efficiency of $\alpha_\mathrm{sc}=100$ and an overshoot parameter of $\alpha_\mathrm{ov} = 0.335$ (left) and $\alpha_\mathrm{ov} = 1.2$ (right) for LMC (top) and MW (bottom) initial compositions, with an initial rotation velocity of $100\,\mathrm{km}\,\mathrm{s}^{-1}$. Initial masses are in the range $4\,\mathrm{M}_\odot \le M_\mathrm{ZAMS} \le 107\,\mathrm{M}_\odot$. The line colour follows the surface Eddington factor. Black crosses show the points in the evolution where core hydrogen burning ends, and red asterisks mark the end of the evolution, when core carbon burning ends. The thick magenta line maps the HD limit.}
\label{fig:HR2}
\end{figure*}

Figure \ref{fig:HR1} shows evolutionary tracks for LMC and MW compositions and an initial rotation velocity of $V_\mathrm{i}=200\,\mathrm{km}\,\mathrm{s}^{-1}$, for $\alpha_\mathrm{sc}=100$ and step overshooting with $\alpha_\mathrm{ov}=0.335$ and $\alpha_\mathrm{ov}=1.2$. Points at intervals of $50\,000\,\mathrm{yr}$ along the evolution are marked with dots, to illustrate the duration of different phases, and the relative expected number of stars at each part of the evolution for each initial mass. It can be seen that for $\alpha_\mathrm{ov}=0.335$ numerous models spend a long time beyond the HD limit, while for $\alpha_\mathrm{ov}=1.2$ the evolutionary tracks almost do not cross the limit, and then only for a brief time if they do. The value of $\alpha_\mathrm{ov}=1.2$ is probably excessive, as a gap is present between the tracks and the diagonal part of the HD limit. We do not suggest that $\alpha_\mathrm{ov}=1.2$ is a reasonable choice, but rather use it to demonstrate the strong effect of overshooting. To test the models a statistical analysis is discussed in Section \ref{sec:popsynvsobs}.

Figure \ref{fig:HR2} shows evolutionary tracks for LMC and MW compositions and an initial rotation velocity of $V_\mathrm{i}=100\,\mathrm{km}\,\mathrm{s}^{-1}$ (two times slower compared to the models in Figure \ref{fig:HR1}), for the same mixing parameters as in Figure \ref{fig:HR1} ($\alpha_\mathrm{sc}=100$, and $\alpha_\mathrm{ov}=0.335$ or $\alpha_\mathrm{ov}=1.2$), with the colour along the tracks showing the ratio between the stellar luminosity and the Eddington luminosity $\Gamma_\mathrm{Edd}$ (computed by \texttt{MESA}, taking into account the gas opacity). Compared to the tracks in Figure \ref{fig:HR1}, the models with $\alpha_\mathrm{ov}=1.2$ in Figure \ref{fig:HR2} evolve somewhat further beyond the HD limit, because of the slightly reduced rotational mixing. We point out in Figure \ref{fig:HR2} the possibility of stellar models to reside beyond the HD limit while not exceeding their Eddington luminosity. Also marked are the terminal age main sequence (TAMS\footnote{The TAMS is defined as the point where the central hydrogen mass fraction drops below $0.01$.}) and final positions on the HRD\footnote{The final position is marked only for models which ignited carbon in their centre, and are therefore considered to be CCSN progenitors.}. For $\alpha_\mathrm{ov}=0.335$, the TAMS location is beyond the HD limit for the highest initial masses. For $\alpha_\mathrm{ov}=1.2$, the TAMS moves redward for the lower masses, but for the higher masses it can move bluewards. There are less pre-SN red supergiants for $\alpha_\mathrm{ov}=1.2$, though for such a large extent of overshooting lower initial masses might give rise to CCSNe whose progenitors are red supergiants.

\begin{figure}
\centering
\begin{tabular}{c}
\includegraphics[width=0.465\textwidth]{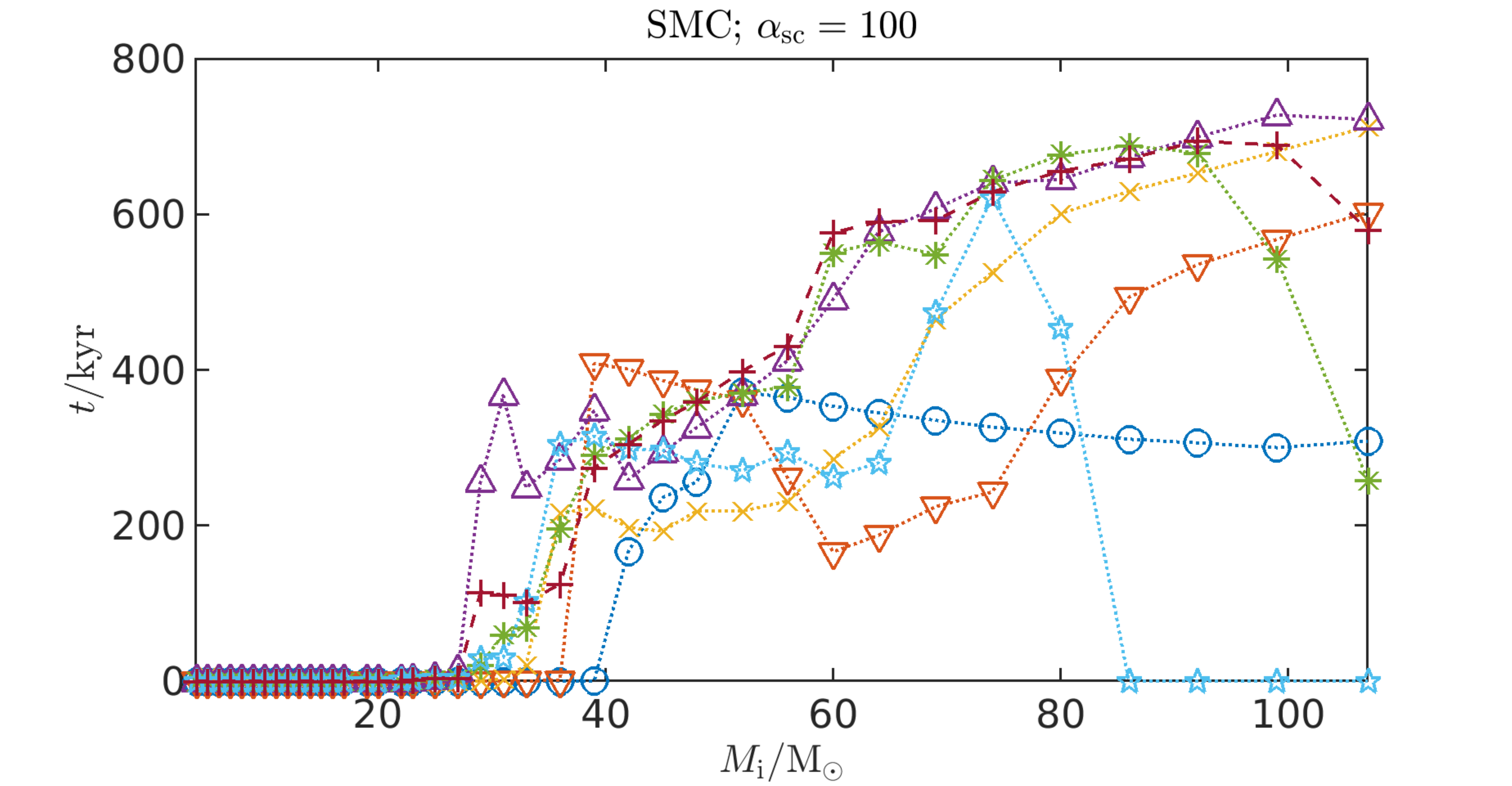} \\ 
\includegraphics[width=0.465\textwidth]{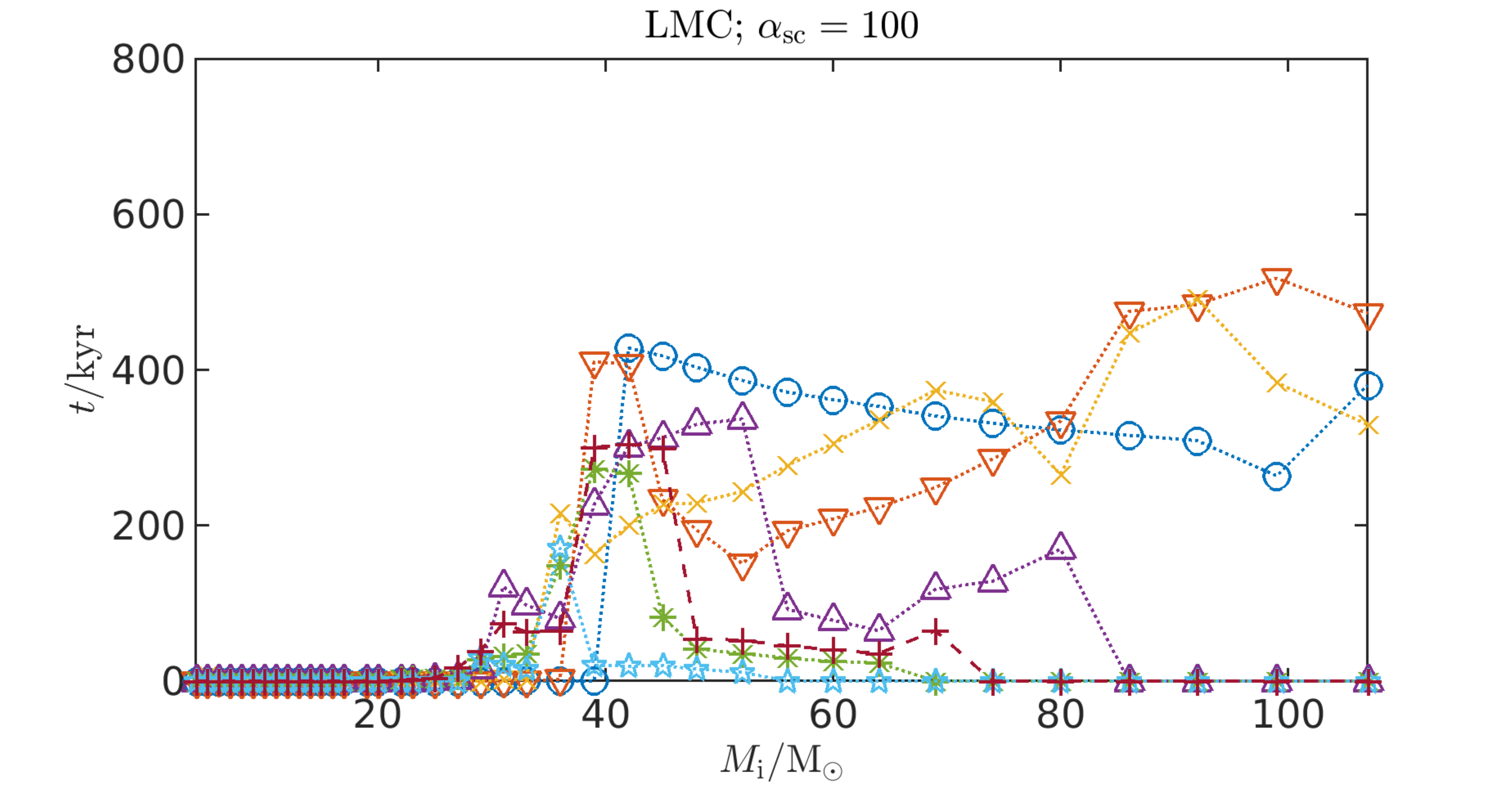} \\ 
\includegraphics[width=0.465\textwidth]{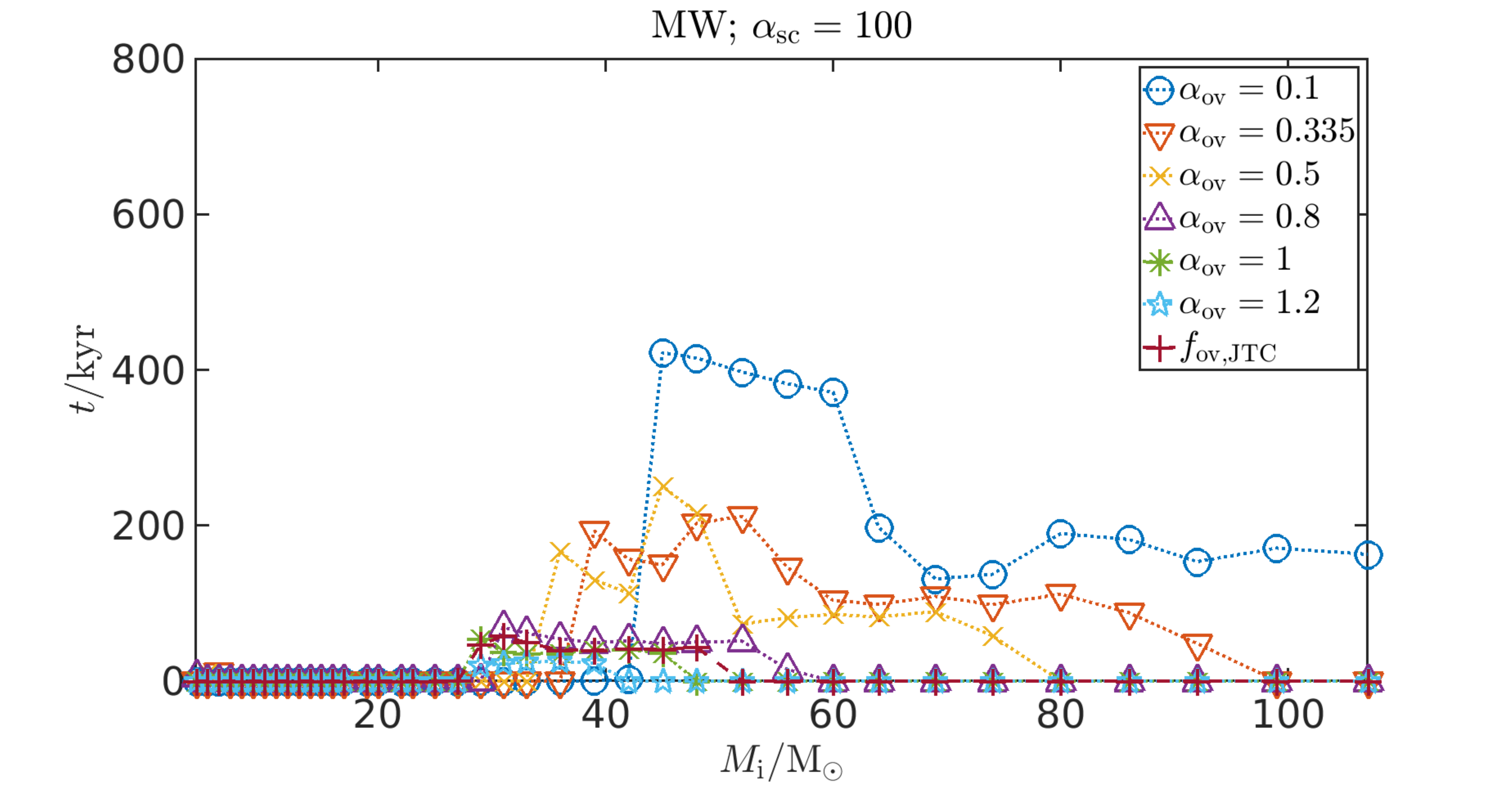} \\ 
\end{tabular}
\caption{Time spent beyond the HD limit as function of ZAMS mass for SMC (top), LMC (middle) and MW (bottom) composition, for different core overshooting prescriptions. For each initial mass, the maximal value out of all initial rotation rates is taken. Points labeled as $f_\mathrm{ov,JTC}$ use the exponential overshooting coefficient $f_\mathrm{ov}$ as function of core properties described by \citealt{JTC}.}
\label{fig:dtHD1}
\end{figure}

The duration of the evolutionary stage when a star is located on the HRD above the HD limit is shown in Figure \ref{fig:dtHD1} for various parameters. For each composition, initial mass, and overshooting scheme, there are $3$ different initial rotation velocities, and we take the longest time from the three tracks. For $\alpha_\mathrm{ov}=0.335$, massive stars spend $\ga 100\, 000 \,\mathrm{yr}$ beyond the HD limit. The highest masses with MW composition do not cross the HD limit thanks to mass removal by winds (besides the models with $\alpha_\mathrm{ov}=0.1$), but as the  rate depends on the metallicity, this effect is not present in the SMC and LMC models, except for those with significantly enhanced overshooting. With increasing $\alpha_\mathrm{ov}$, the LMC and MW models spend less time beyond the HD limit.
    
For the SMC the models with $30 \la M_\mathrm{ZAMS} / \mathrm{M}_\odot \la 80$ always cross the HD limit and spend a long time beyond it. This is because crossing the HD limit is a result of the combination of mixing and mass loss. With increased mixing, envelope material is used as fuel in the core, both increasing the core mass and luminosity and therefore the mass loss, while at the same time decreasing the envelope mass which needs to be removed by winds.

\begin{figure}
\centering
\includegraphics[width=0.465\textwidth]{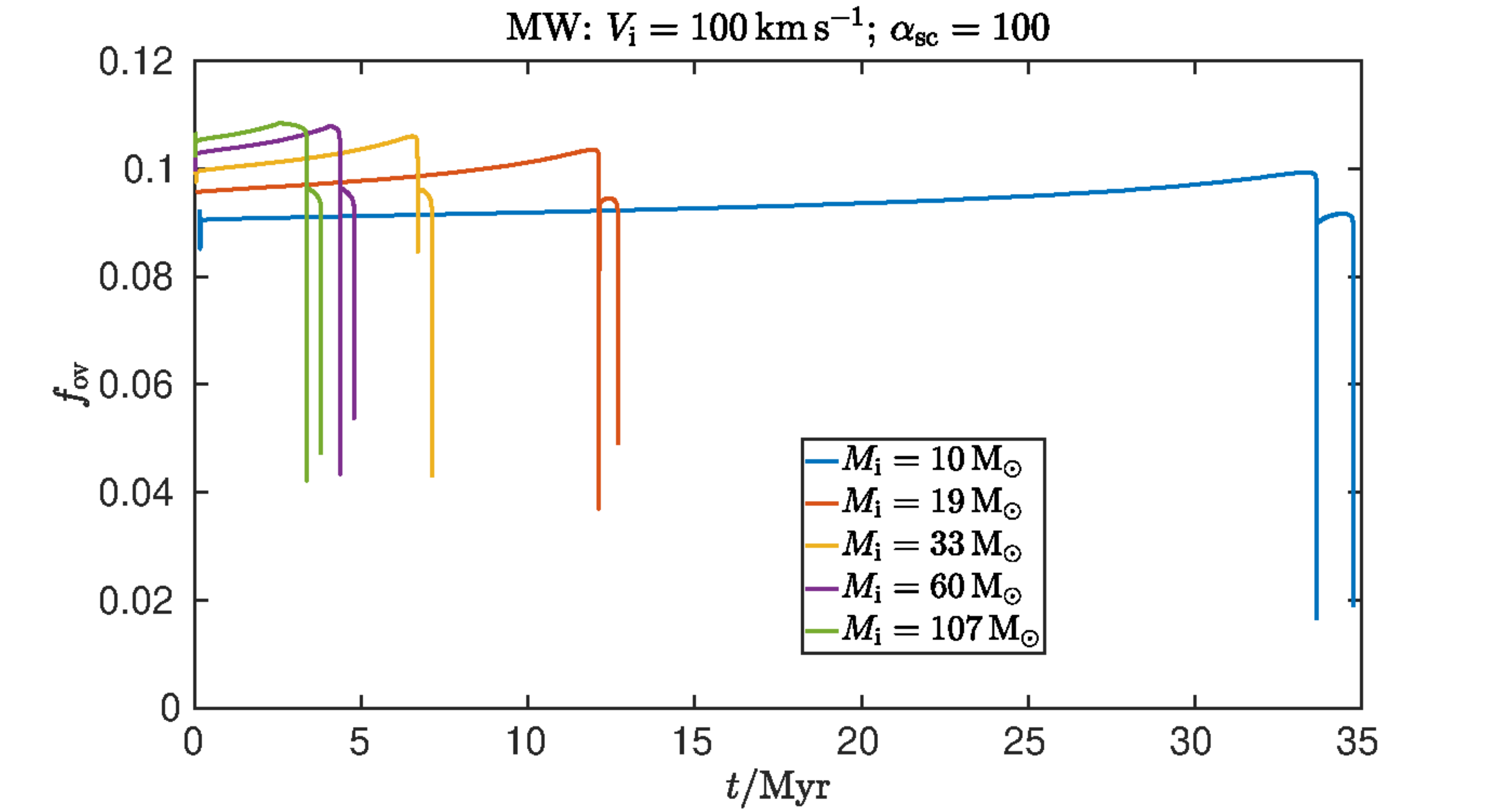} \\ 
\caption{Exponential decay scale as fraction of the pressure scale height as function of time for several models with MW composition and an initial rotation velocity of $V_\mathrm{i} = 100\,\mathrm{km}\,\mathrm{s}^{-1}$ as calculated by using equation (\ref{eq:fov}).}
\label{fig:fov}
\end{figure}

The models which use the overshooting prescription of \cite{JTC} give quantitatively similar results to those with $\alpha_\mathrm{ov}=1$, with $f_\mathrm{ov}$ varying slightly with time and initial mass but generally close to $f_\mathrm{ov}\approx 0.1$ as shown in Figure \ref{fig:fov}. \cite{JTC} motivate their prescription as describing enhanced rotational mixing caused by core anisotropy. In that sense there is a physical motivation for enhanced mixing. However, \cite{JTC} caution that their prescription appears to overestimate $f_\mathrm{ov}$ relative to what \cite{ClaretTorres2017} infer from observations. Furthermore, while the usage of $f_\mathrm{ov}$ is convenient, their mixing mechanism does not exhibit an exponentially decaying geometry like the $f_\mathrm{ov}$ prescription assumes and so this implementation does not accurately describe the physics. Our results imply that there is a strong motivation to further investigate the mixing mechanism proposed by \cite{JTC}.
    
\section{Statistical comparisons between synthetic populations and observations}
\label{sec:popsynvsobs}

\subsection{Our sample compilation}
\label{subsec:catalogue}

We compile the luminosities and temperatures of all known CSGs in the Magellanic Clouds with a luminosity of \mbox{$\log_{10} \left( L / \mathrm{L}_\odot\right) = 4.7$} or higher. Unlike \citet{HDrevisited}, who only considered RSGs, we consider RSGs, YSGs, and cool BSGs. The reason is that the HD limit stretches over a large region on the HRD populated by these various spectral types. For simplicity, we only consider here the horizontal part of the HD limit, which extends up to effective temperatures of $\approx 12.5\,\mathrm{kK}$. Following standard convention \cite[e.g.,][]{Drout2012}, we define RSGs, YSGs, and cool BSGs in the temperature regimes $T_{\rm eff} \le 4\,800\,$K,  $4\,800\,{\rm K} < T_{\rm eff} < 7\,500\,{\rm K}$, and $7\,500 \le T_{\rm eff} \le 12\,500\,{\rm K}$, respectively.

For the LMC, we cross-match the RSG list of \citet{HDrevisited} with the RSG-YSG list of \citet{Neugent2012}. For targets that appear in both compilations, we adopt temperatures and luminosities from \citet{HDrevisited}, who derived these parameters from a complete spectral energy distribution (SED) fitting. When temperatures are not specified, we use calibrations between spectral types and temperatures by \citet{Tabernero2018} to derive the temperature. Since both studies claim to be complete for $\log_{10}\left(L / \mathrm{L}_\odot \right) \ge 4.7$, we only include objects exceeding this threshold. 

The catalogue of \citet{Neugent2012} also includes  stars hotter than $7\,500\,\mathrm{K}$, which are considered BSGs. Since their study is not necessarily complete for BSGs, we extended our catalogue by retrieving the  Gaia DR2 catalogue \citep{GaiaDR2} centered on the LMC with a search radius of $5.5$ degrees. To identify cool BSG candidates, we filtered all LMC stars in the Gaia DR2 catalogue with Gaia fluxes fulfilling the criteria $-0.1 < Bp-Rp < 0.6\,$mag and  $G < 15.5\,$mag, accounting for a typical reddening value of $E_{B-V} = 0.09$ \citep[e.g.][]{Fitzpatrick1990}. We then cross-matched our list with the SIMBAD catalogue to retrieve spectral types using the {\sc Vizier X-match} service. Main references are \citet{Sanduleak1970}, \citet{Ardeberg1972}, \citet{Stock1976}, \citet{Evans2006}, and \citet{urbaneja_lmc_2017}. All identified targets were classified before and have spectral types consistent with BSGs, and the majority of those were included in previous spectroscopic analyses of CSGs in the LMC. 

All stars with spectral types earlier than B7 in our final list were removed, including a few WR stars. Finally We included four LBVs from the compilation given by \citet{Smith2019}. 

For all remaining objects, we extracted radial velocities (RVs) and proper motions (PMs) from the SIMBAD database \citep{Wenger2000}. The sources of the RVs were predominantly the Gaia DR2 catalogue \citep{GaiaDR2},  \citet{MasseyOlsen2003}, \citet{Neugent2012}, \citet{Fehrenbach1972}, and \citet{Fehrenbach1982}. PMs originate in the Gaia DR2 catalogue for all sources but ten, for which they are retrieved from Gaia DR1 \citep{GaiaDR1}. The mean PM is $1.78\,{\rm mas}\,{\rm yr}^{-1}$ with a standard deviation of $0.3\,{\rm mas}\,{\rm yr}^{-1}$, which reflects the measurement limit of Gaia. There are $15$ outliers with PMs larger than $4\,{\rm mas}\,{\rm yr}^{-1}$ within their respective errors, and they are omitted from our sample to ensure that we do not include foreground Galactic objects. 

When available, $T_{\rm eff}$ and $\log L$ values for the cool BSGs were adopted from  \citet{urbaneja_lmc_2017} and \citet{Smith2019}. Otherwise, we used spectral-type calibrations by \citet{Fitzpatrick1990} to derive the effective temperatures and estimated the extinction parameters based on the expected intrinsic colours. We then used bolometric corrections following  \citet{Flower1996} and \citet{Torres2010}, assuming a distance of $49.97\,\mathrm{kpc}$ \citep{Pietrzynski2013}. The final list for the LMC comprises $375$ stars: $265$ RSGs, $39$ YSGs, and $71$ cool BSGs (four of which are LBVs).

For the SMC, we repeat this procedure using the RSGs listed by \citet{HDrevisited} and the YSGs listed by \citet{Neugent2010}. The RVs, PMs, and spectral types are again extracted using SIMBAD. The RVs and PMs originate predominantly from the Gaia DR2 catalogue, but also from \citet{MasseyOlsen2003},  \citet{Neugent2010}, and  \citet{GonzalesFernandez2015}. The spectral types are retrieved from  \citet{Feast1960}, \citet{Dubois1977}, \citet{Humphreys1983}, \citet{Lennon1997}, and \citet{Dufton2000}. Again, all objects earlier than B7 are removed. We identify about ten outliers with $\mathrm{PM} > 4\,{\rm mas}\,{\rm yr}^{-1}$, which are removed from our sample.  When available, $T_{\rm eff}$ and $\log L$ values for the cool BSGs were adopted from  \citet{Dufton2000}. Otherwise, we use calibrations by \citet{Evans2003} to derive the effective temperatures, make the same assumptions as \citet{Neugent2010} regarding the reddening and the distance towards the SMC, and use the same relations as above to derive the bolometric corrections and luminosities. The final list comprises $179$ stars: $140$ RSGs, $7$ YSGs, and $32$ cool BSGs.
 
We note that accurate derivations of $T_{\rm eff}$ and $\log L$ should rely on the fitting of SEDs. However, given the statistical nature of our study, the calibrations used above should be sufficient  for our purpose \citep[e.g.][]{Neugent2010, Neugent2012}.

\subsection{Population synthesis}
\label{subsec:popsyn}

We construct synthetic populations by generating random initial masses according to a Salpeter IMF. The initial rotation velocity is chosen according to the observed distributions for the LMC \citep{ramachandran_stellar_2018b} and the SMC \citep{ramachandran_testing_2019}. For both the initial mass and velocity, the nearest values available in our models are used to chose a stellar evolution track to follow, rather than interpolating between tracks. This results in $58\% $ of SMC models being assigned an initial rotation velocity of $100\,\mathrm{km}\,\mathrm{s}^{-1}$, $31\% $ getting $200\,\mathrm{km}\,\mathrm{s}^{-1}$ and $11\% $ with $300\,\mathrm{km}\,\mathrm{s}^{-1}$. For the LMC the corresponding percentages are $80\% $, $19\% $ and $1\% $.

The stellar age is chosen according to a uniform distribution, corresponding to a constant star-formation rate (SFR)\footnote{This is a reasonable assumption as we are interested in young stars, though in general the SFR is not constant in the SMC and the LMC (Section \ref{subsec:SFH}).}. The stellar properties are interpolated from the evolutionary track according to the generated stellar age. If the generated age is longer than the lifetime of the computed stellar evolution track, the star is discarded.

\begin{figure}
\centering
\includegraphics[width=0.495\textwidth]{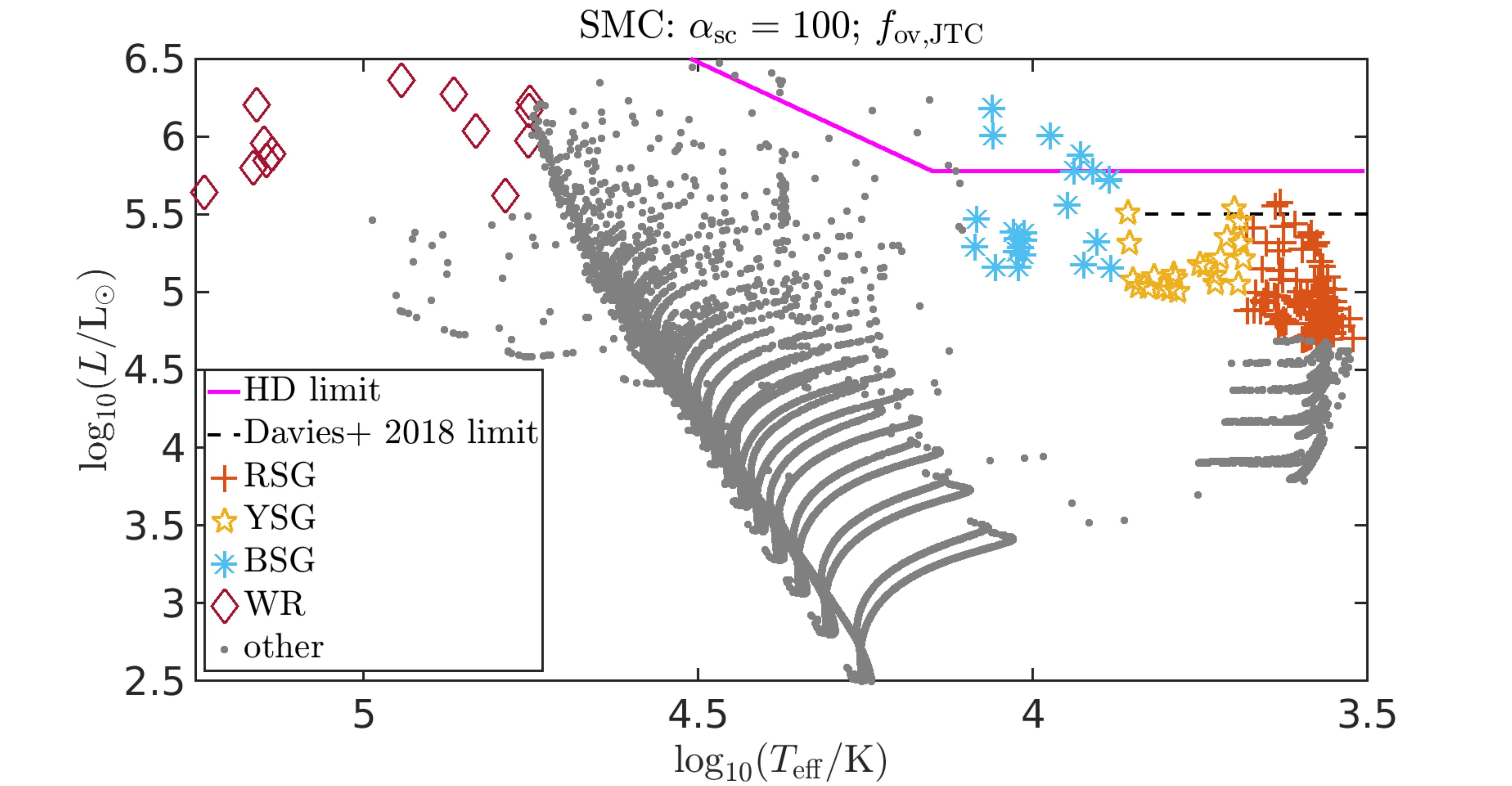} \\
\includegraphics[width=0.495\textwidth]{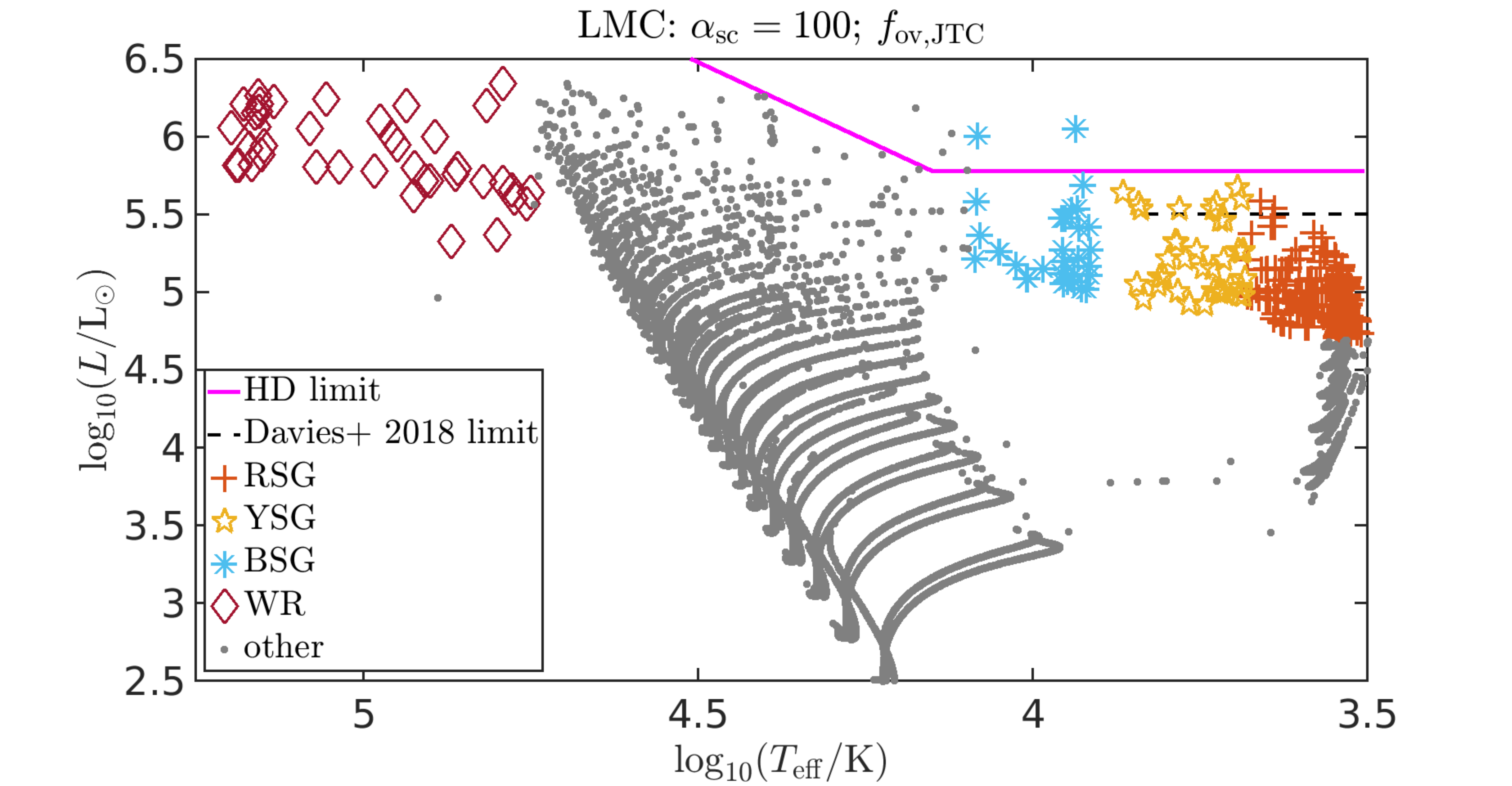} \\
\caption{Example synthetic populations of stars with SMC (top) and LMC (bottom) initial compositions.
}
\label{fig:popsynexample}
\end{figure}

Random stars are generated until the combined number of RSGs and YSGs ($T_\mathrm{eff}<7500\,\mathrm{K}$) in the luminosity interval $4.7 \le \log_{10} (L/ \mathrm{L}_\odot) < 5.2$ matches the observed number -- this is our normalisation. For the SMC this gives $134$ red and yellow supergiants, and for the LMC $274$. An example of a synthetic population is presented in Figure \ref{fig:popsynexample}.

\subsection{Observations vs. simulations}
\label{subsec:obsvscalc}

\begin{figure*}
\centering
\begin{tabular}{cc}
\includegraphics[width=0.465\textwidth]{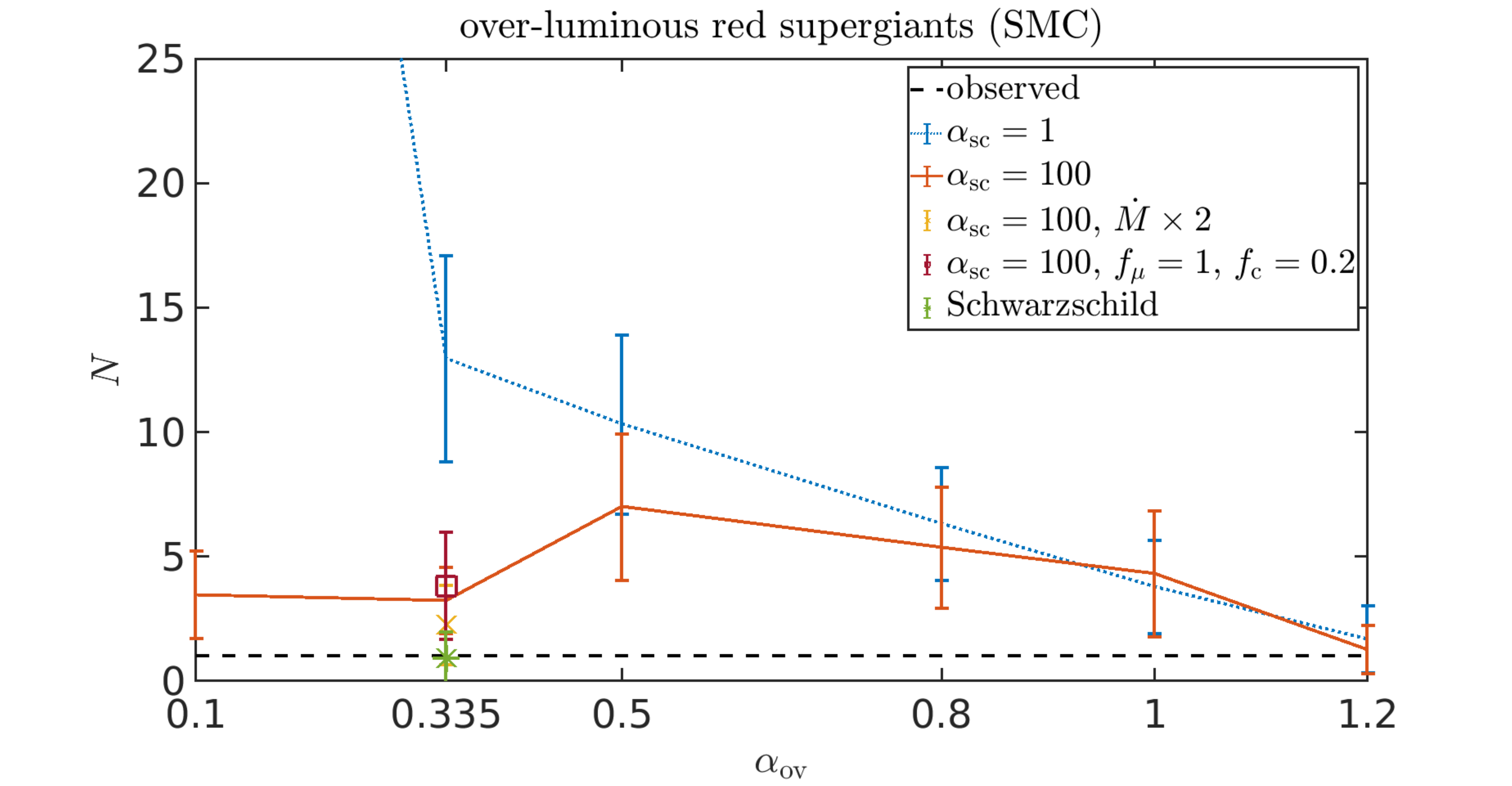} &
\includegraphics[width=0.465\textwidth]{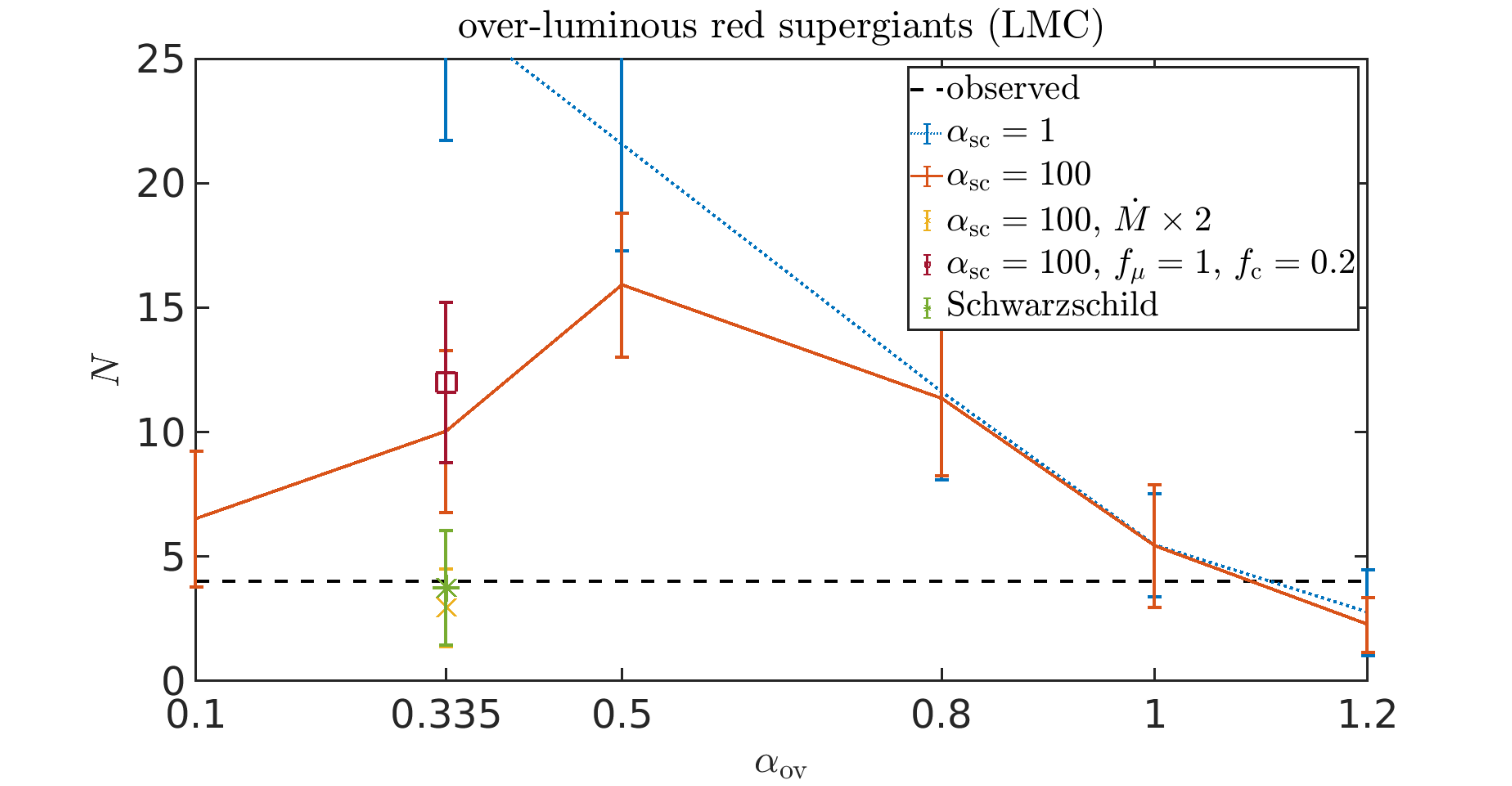} \\ 
\includegraphics[width=0.465\textwidth]{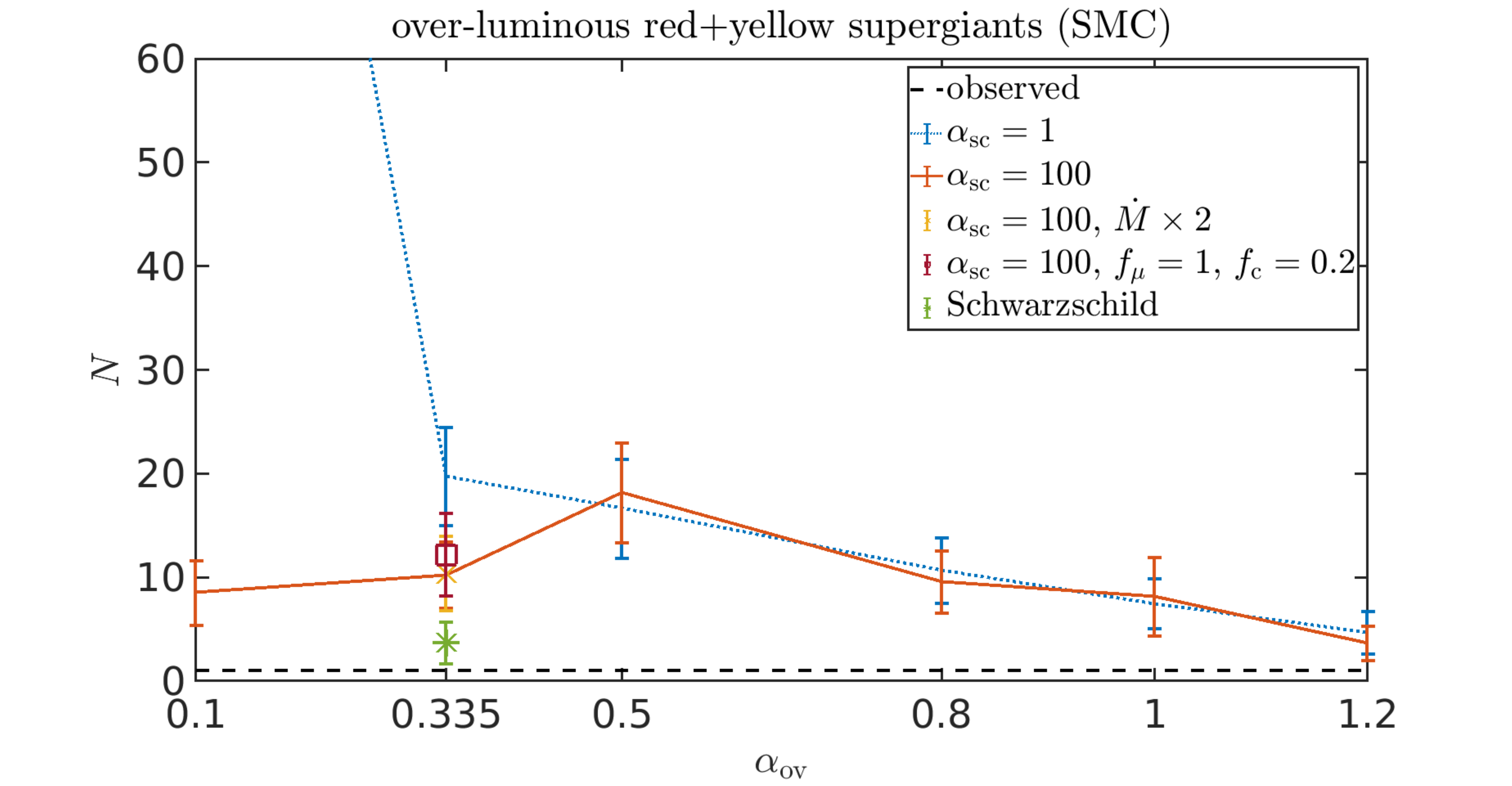} & 
\includegraphics[width=0.465\textwidth]{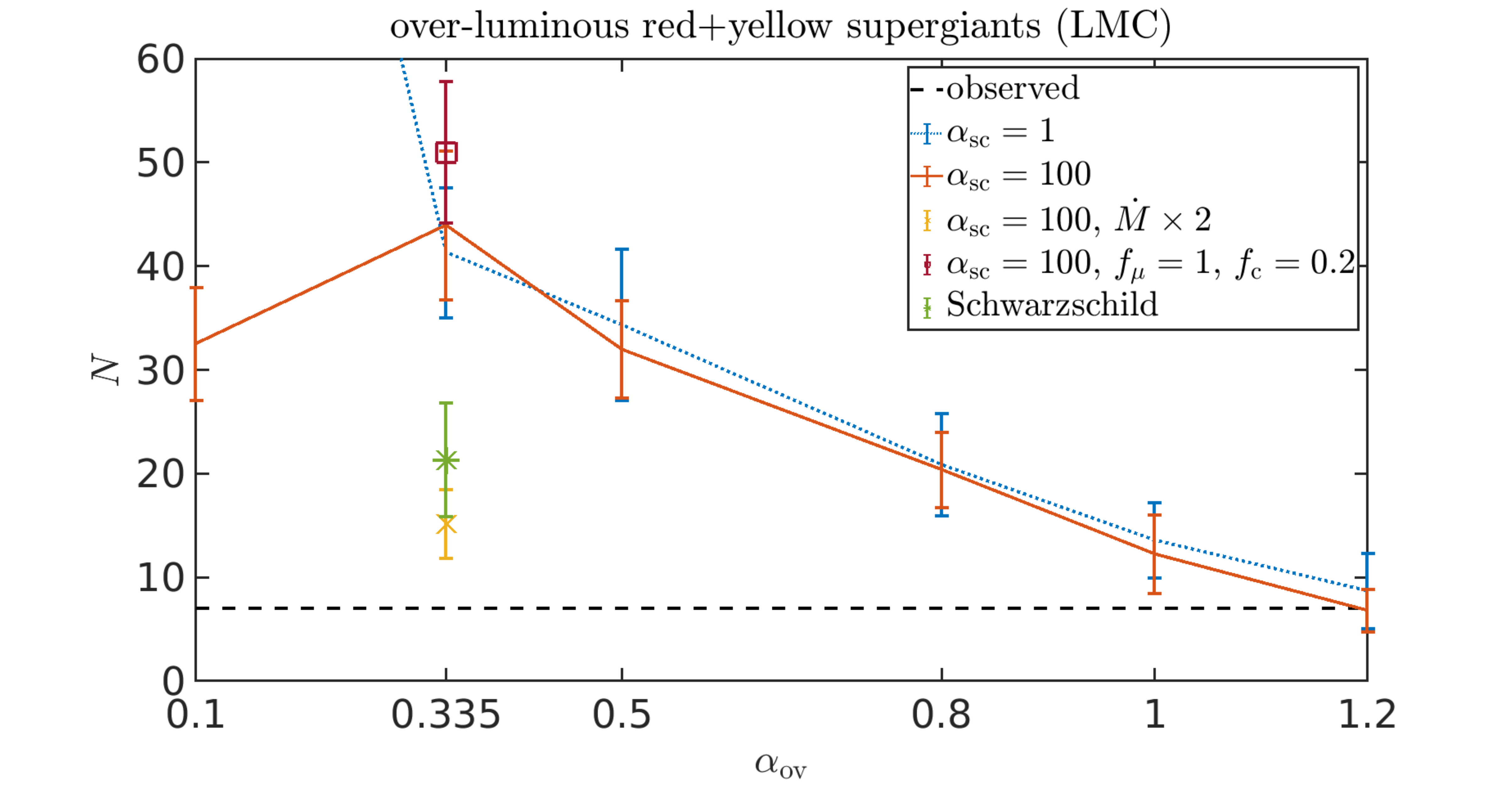} \\ 
\includegraphics[width=0.465\textwidth]{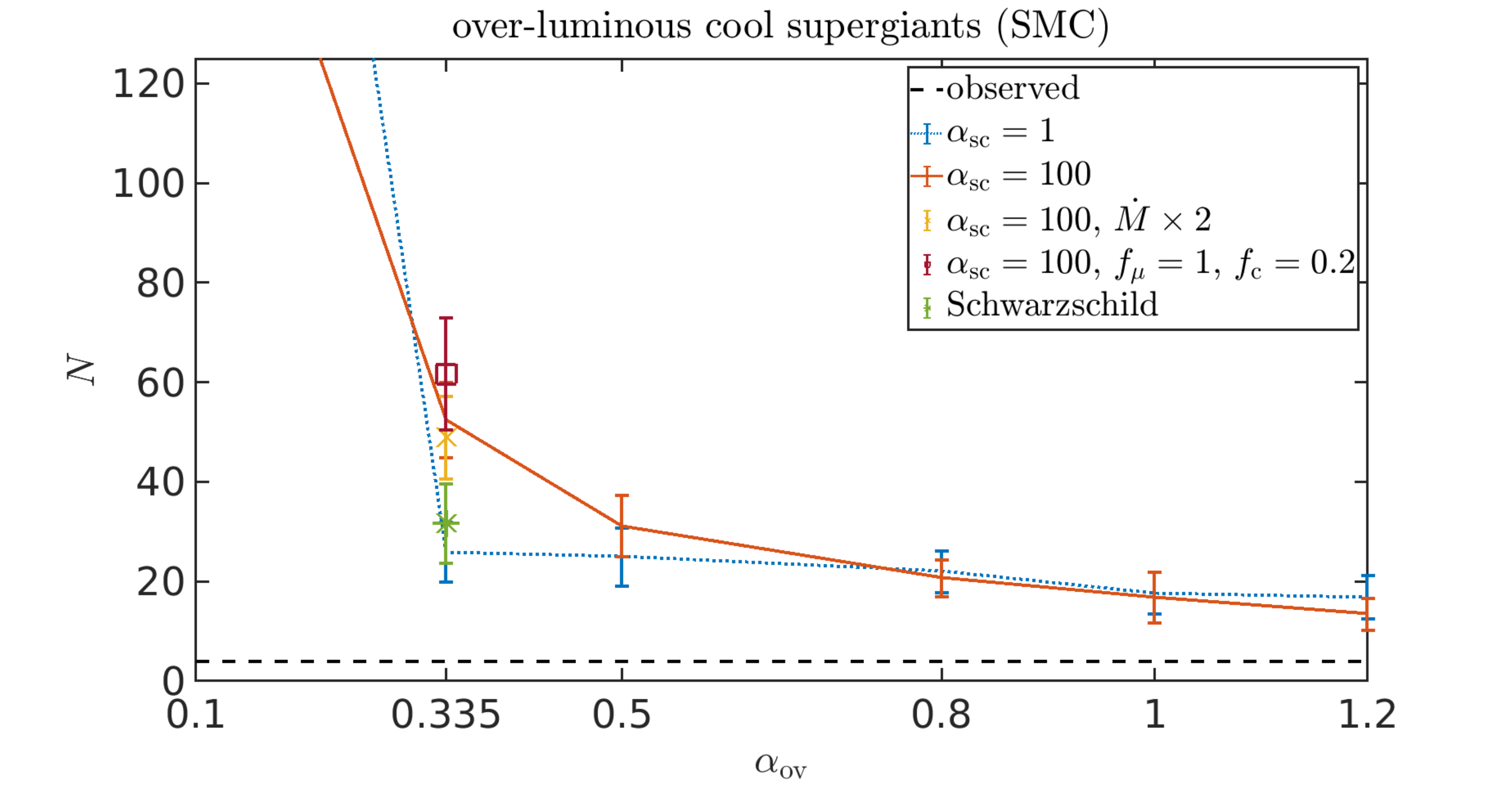} &
\includegraphics[width=0.465\textwidth]{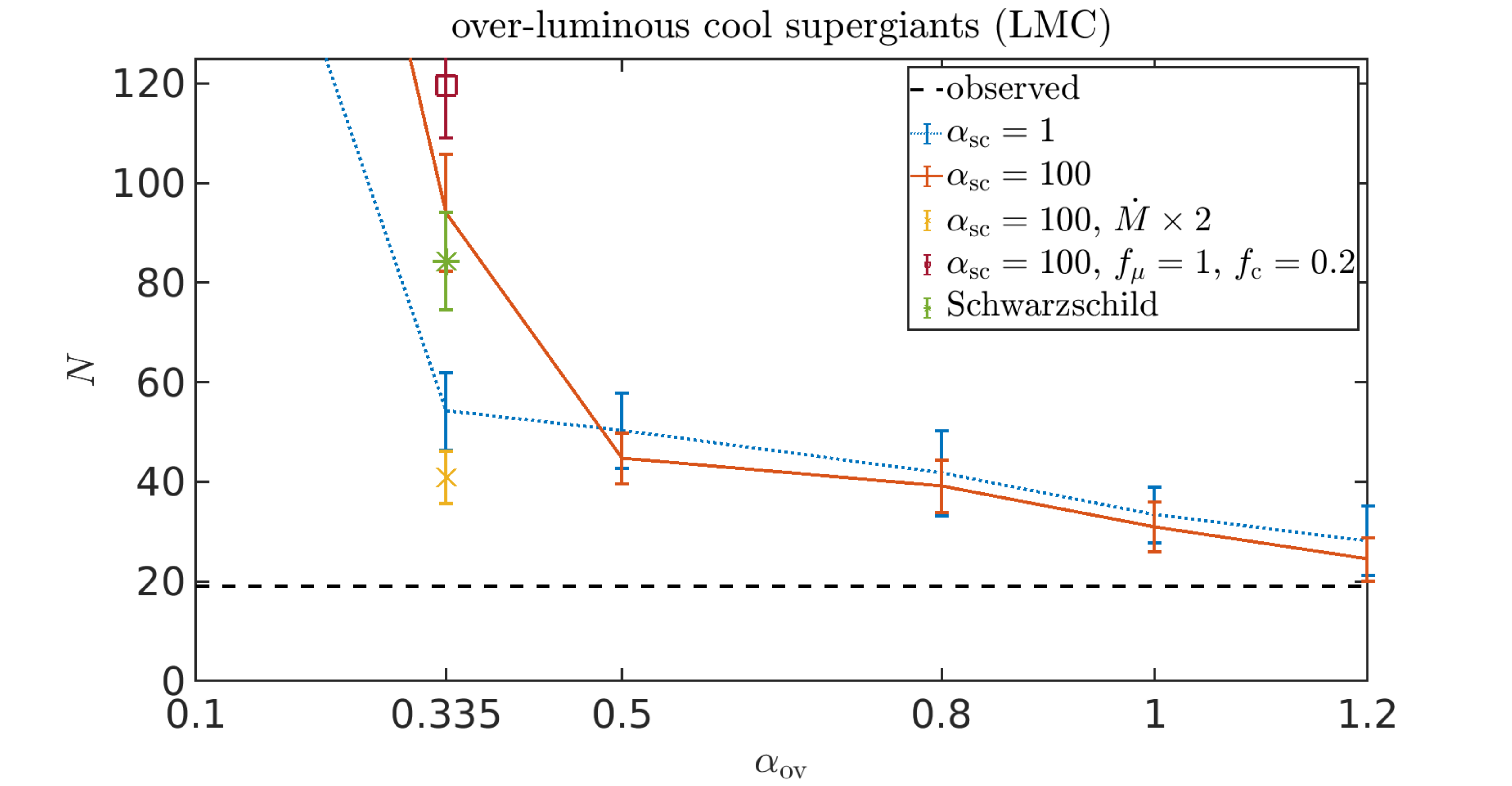} \\ 
\end{tabular}
\caption{Number of supergiant stars with $\log_{10}\left(L / \mathrm{L}_\odot\right) \ge 5.4$ in synthetic populations as function of the overshooting extent, for two values of the semiconvective mixing efficiency, as well as models employing the Schwarzschild stability criterion instead of Ledoux, for the SMC (left) and the LMC (right). For $ \alpha_\mathrm{sc}=100 $, models with doubled mass loss or alternative parameters for rotational mixing are also shown. The top panels show the number of over-luminous red ($T_\mathrm{eff} \le 4\, 800 \, \mathrm{K} $) supergiant stars, the middle panels show red and yellow ($T_\mathrm{eff} < 7\, 500 \, \mathrm{K} $) supergiant stars, and the bottom panels show cool ($T_\mathrm{eff} \le 12\, 500 \, \mathrm{K} $) supergiant stars.}
\label{fig:popsyn1vs100}
\end{figure*}

For each of the $17$ sets of modelling assumptions (Table \ref{tab:modelassumptions}) we generate $25$ random realisations of such populations to get an error estimate for the computed numbers. We count the number of stars generated which are over-luminous, i.e. those with $\log_{10}\left(L / \mathrm{L}_\odot\right) \ge 5.4$. The results for the $850$ synthetic populations are summarised in Table \ref{tab:sgex} and Table \ref{tab:sgexx}, and in Figure \ref{fig:popsyn1vs100} we show the number of over-luminous supergiants in our synthetic populations for all models which employ step overshooting. Tables \ref{tab:sgex} and  \ref{tab:sgexx} show that looking only at red supergiants leads to underestimating the excess of over-luminous stellar models. The excess generally decreases with increasing $\alpha_\mathrm{ov}$, although for efficient semiconvective mixing ($\alpha_\mathrm{sc}=100$) the behaviour is non-monotonic for the red and yellow regimes. The models employing the overshooting prescription given by equation (\ref{eq:fov}) produce very similar results to those with $\alpha_\mathrm{ov}=1$.
\begin{table}
\centering
\caption{Excess of over-luminous stars in synthetic populations.}
\begin{threeparttable}
\begin{tabular}{c|cc|ccc|cc}
\hline
\hline
& overshooting & $\alpha_\mathrm{sc}$ & $N_\mathrm{excess,RSG+YSG}$ & $N_\mathrm{excess,RSG}$ \\
\hline
SMC & $\alpha_\mathrm{ov}=0.1$ & $1$ & $152 \pm 15$ & $79 \pm 10$ \\
SMC & $\alpha_\mathrm{ov}=0.335$ & $1$ & $20 \pm 5$ & $13 \pm 4$ \\
SMC & $\alpha_\mathrm{ov}=0.5$ & $1$ & $17 \pm 5$ & $10 \pm 4$ \\
SMC & $\alpha_\mathrm{ov}=0.8$ & $1$ & $11 \pm 3$ & $6 \pm 2$ \\
SMC & $\alpha_\mathrm{ov}=1.0$ & $1$ & $7 \pm 2$ & $4 \pm 2$ \\
SMC & $\alpha_\mathrm{ov}=1.2$ & $1$ & $5 \pm 2$ & $2 \pm 1$ \\
SMC & $f_\mathrm{ov,JTC}$ & $1$ & $8 \pm 3$ & $4 \pm 3$ \\
\hline
SMC & $\alpha_\mathrm{ov}=0.1$ & $100$ & $9 \pm 3$ & $3 \pm 2$ \\
SMC & $\alpha_\mathrm{ov}=0.335$ & $100$ & $10 \pm 3$ & $3 \pm 1$ \\
SMC & $\alpha_\mathrm{ov}=0.5$ & $100$ & $18 \pm 5$ & $7 \pm 3$ \\
SMC & $\alpha_\mathrm{ov}=0.8$ & $100$ & $10 \pm 3$ & $5 \pm 2$ \\
SMC & $\alpha_\mathrm{ov}=1.0$ & $100$ & $8 \pm 4$ & $4 \pm 3$ \\
SMC & $\alpha_\mathrm{ov}=1.2$ & $100$ & $4 \pm 2$ & $1 \pm 1$ \\
SMC & $f_\mathrm{ov,JTC}$ & $100$ & $7 \pm 2$ & $4 \pm 2$ \\
\hline
LMC & $\alpha_\mathrm{ov}=0.1$ & $1$ & $145 \pm 15$ & $103 \pm 12$ \\
LMC & $\alpha_\mathrm{ov}=0.335$ & $1$ & $41 \pm 6$ & $27 \pm 5$ \\
LMC & $\alpha_\mathrm{ov}=0.5$ & $1$ & $34 \pm 7$ & $22 \pm 4$ \\
LMC & $\alpha_\mathrm{ov}=0.8$ & $1$ & $21 \pm 5$ & $12 \pm 4$ \\
LMC & $\alpha_\mathrm{ov}=1.0$ & $1$ & $14 \pm 4$ & $5 \pm 2$ \\
LMC & $\alpha_\mathrm{ov}=1.2$ & $1$ & $9 \pm 4$ & $3 \pm 2$ \\
LMC & $f_\mathrm{ov,JTC}$ & $1$ & $14 \pm 3$ & $6 \pm 3$ \\
\hline
LMC & $\alpha_\mathrm{ov}=0.1$ & $100$ & $33 \pm 5$ & $7 \pm 3$ \\
LMC & $\alpha_\mathrm{ov}=0.335$ & $100$ & $44 \pm 7$ & $10 \pm 3$ \\
LMC & $\alpha_\mathrm{ov}=0.5$ & $100$ & $32 \pm 5$ & $16 \pm 3$ \\
LMC & $\alpha_\mathrm{ov}=0.8$ & $100$ & $20 \pm 4$ & $11 \pm 3$ \\
LMC & $\alpha_\mathrm{ov}=1.0$ & $100$ & $12 \pm 4$ & $5 \pm 2$ \\
LMC & $\alpha_\mathrm{ov}=1.2$ & $100$ & $7 \pm 2$ & $2 \pm 1$ \\
LMC & $f_\mathrm{ov,JTC}$ & $100$ & $14 \pm 3$ & $7 \pm 3$ \\
\hline
\end{tabular}
\footnotesize
\end{threeparttable}
\label{tab:sgex}
\end{table}
\begin{table}
\centering
\caption{Excess of over-luminous stars in additional synthetic populations.}
\begin{threeparttable}
\begin{tabular}{cccc}
\hline
\hline
& model assumptions & $N_\mathrm{excess,RSG+YSG}$ & $N_\mathrm{excess,RSG}$ \\
\hline
SMC & Schwarzschild & $4 \pm 2$ & $1 \pm 1$ \\
SMC & $f_c = 0.2$, $f_\mu=1$ & $12 \pm 4$ & $4 \pm 2$ \\
SMC & $\eta_\mathrm{w}=2$ & $10 \pm 4$ & $2 \pm 2$ \\
\hline
LMC & Schwarzschild & $21 \pm 5$ & $4 \pm 2$ \\
LMC & $f_c = 0.2$, $f_\mu=1$ & $51 \pm 7$ & $12 \pm 3$ \\
LMC & $\eta_\mathrm{w}=2$ & $15 \pm 3$ & $3 \pm 2$ \\
\hline
\end{tabular}
\footnotesize
\end{threeparttable}
\label{tab:sgexx}
\end{table}

\begin{figure*}
\centering
\begin{tabular}{cc}
\includegraphics[width=0.465\textwidth]{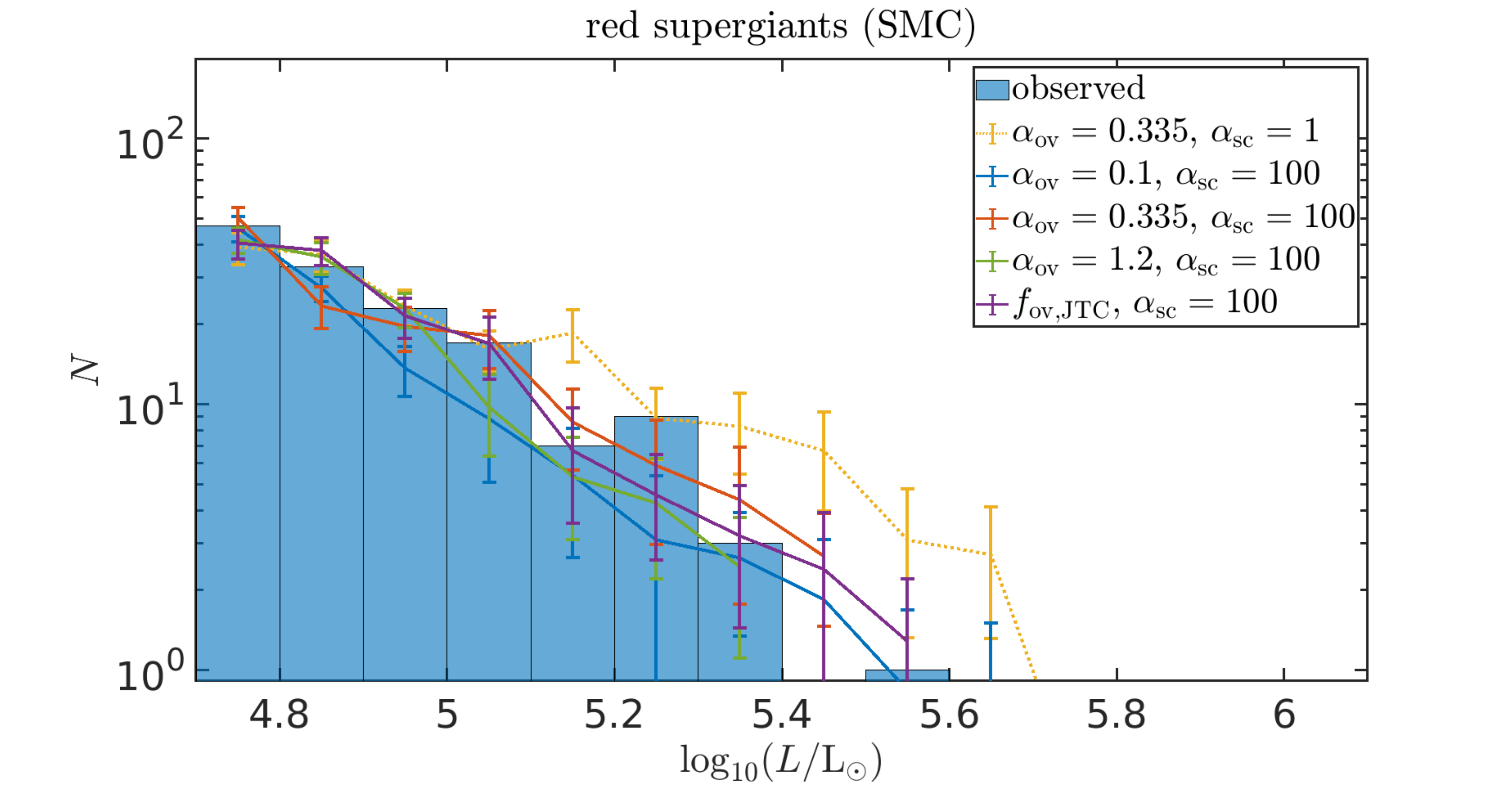} &
\includegraphics[width=0.465\textwidth]{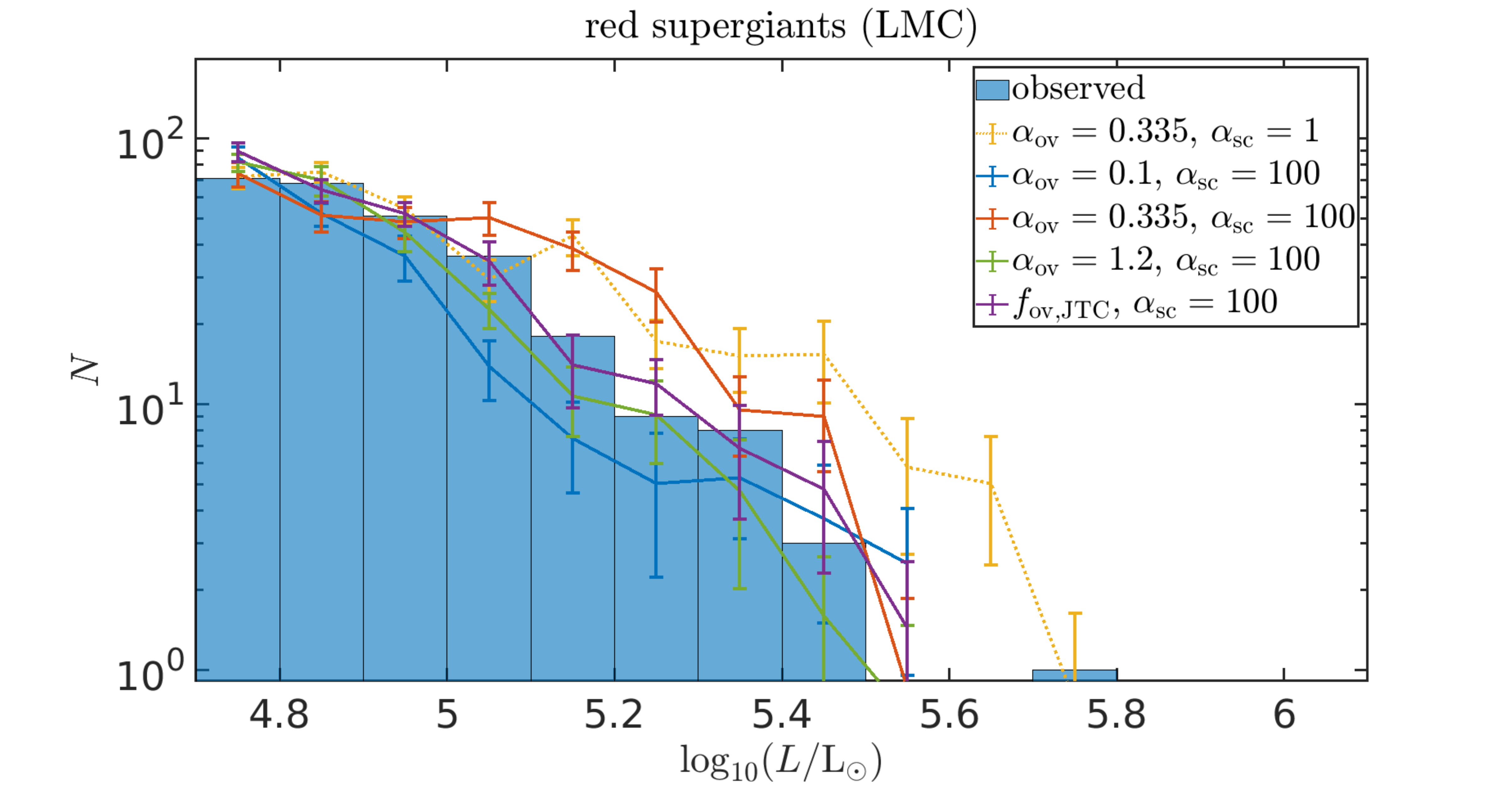} \\ 
\includegraphics[width=0.465\textwidth]{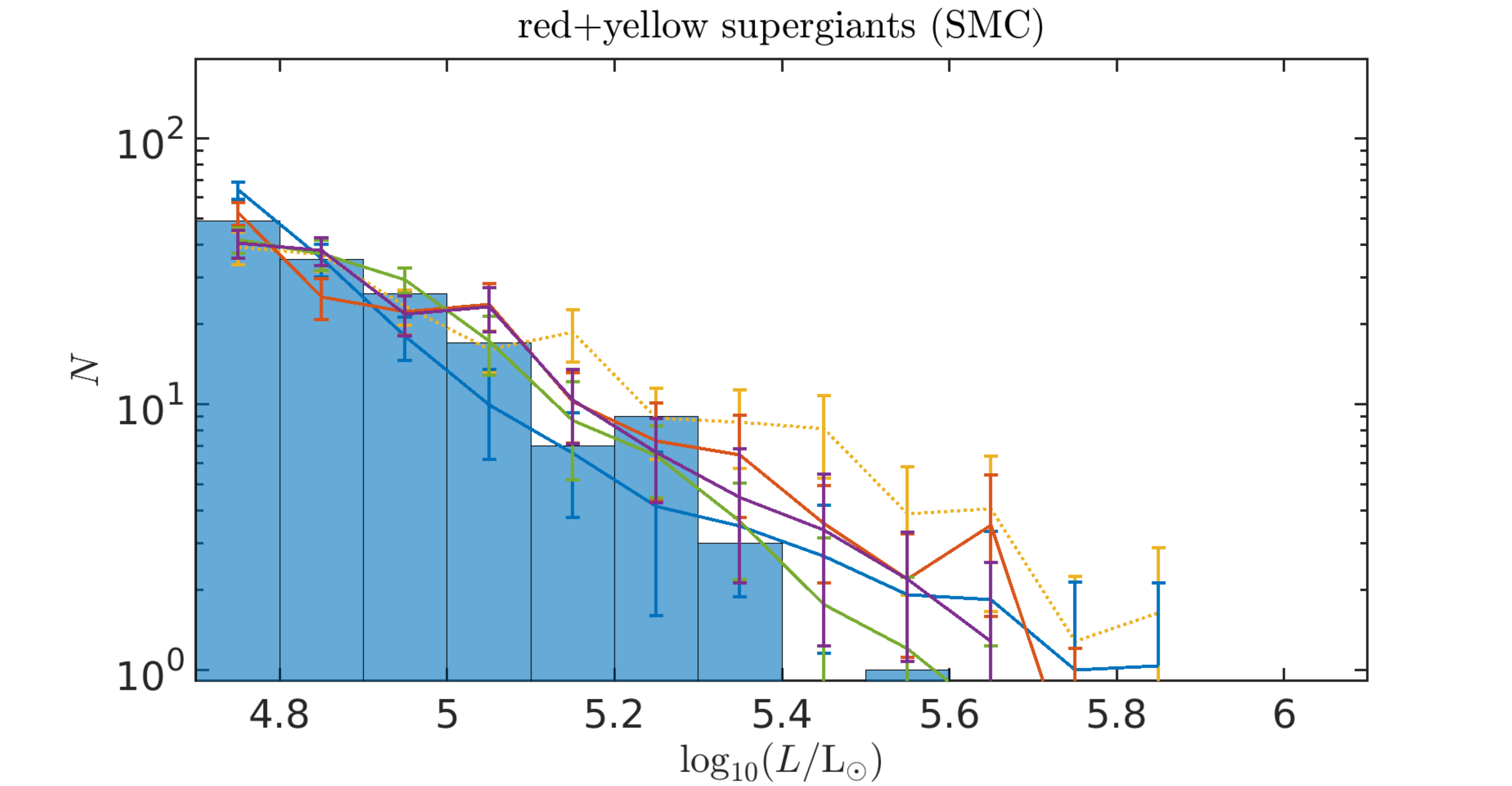} & 
\includegraphics[width=0.465\textwidth]{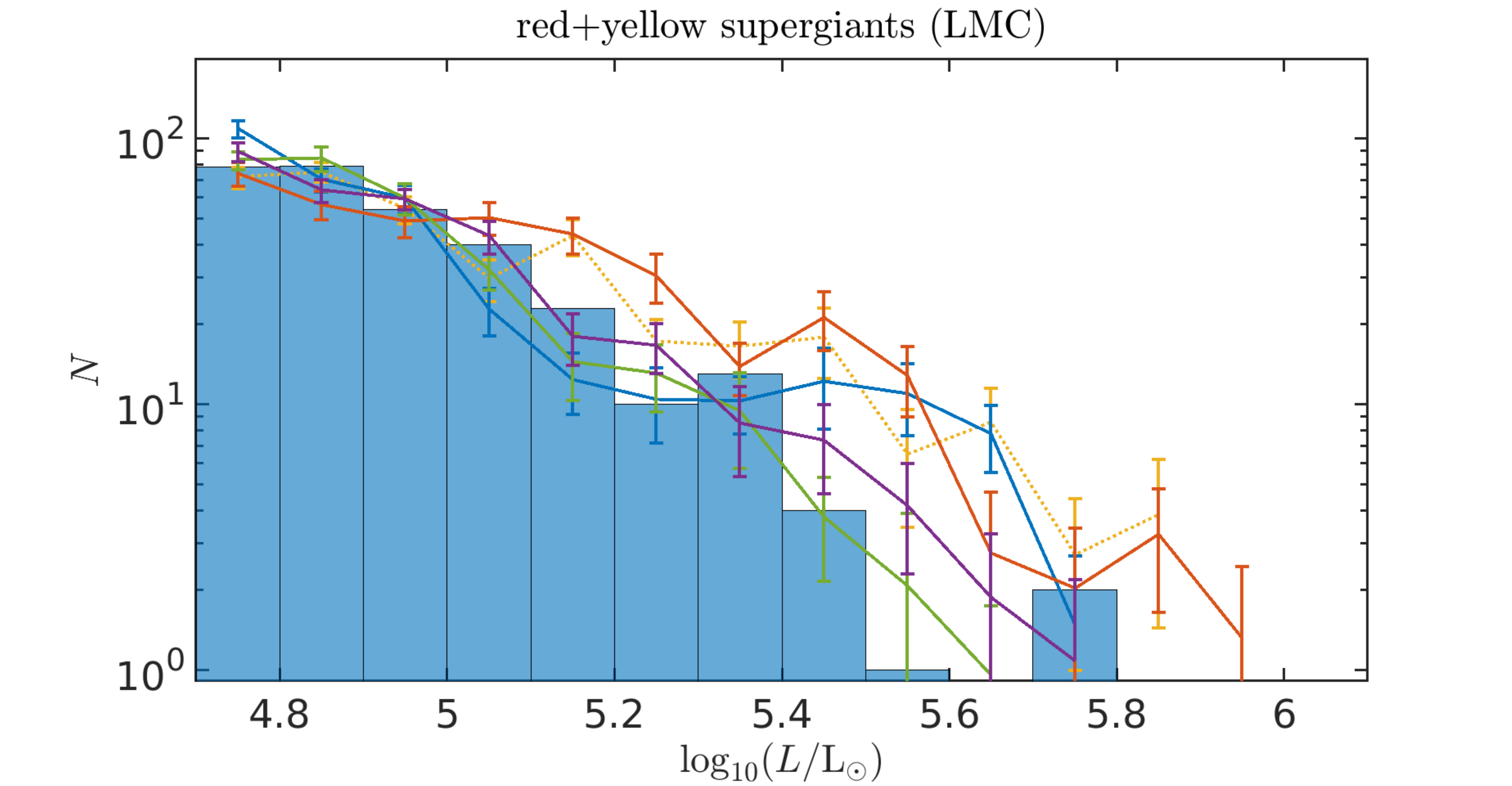} \\ 
\includegraphics[width=0.465\textwidth]{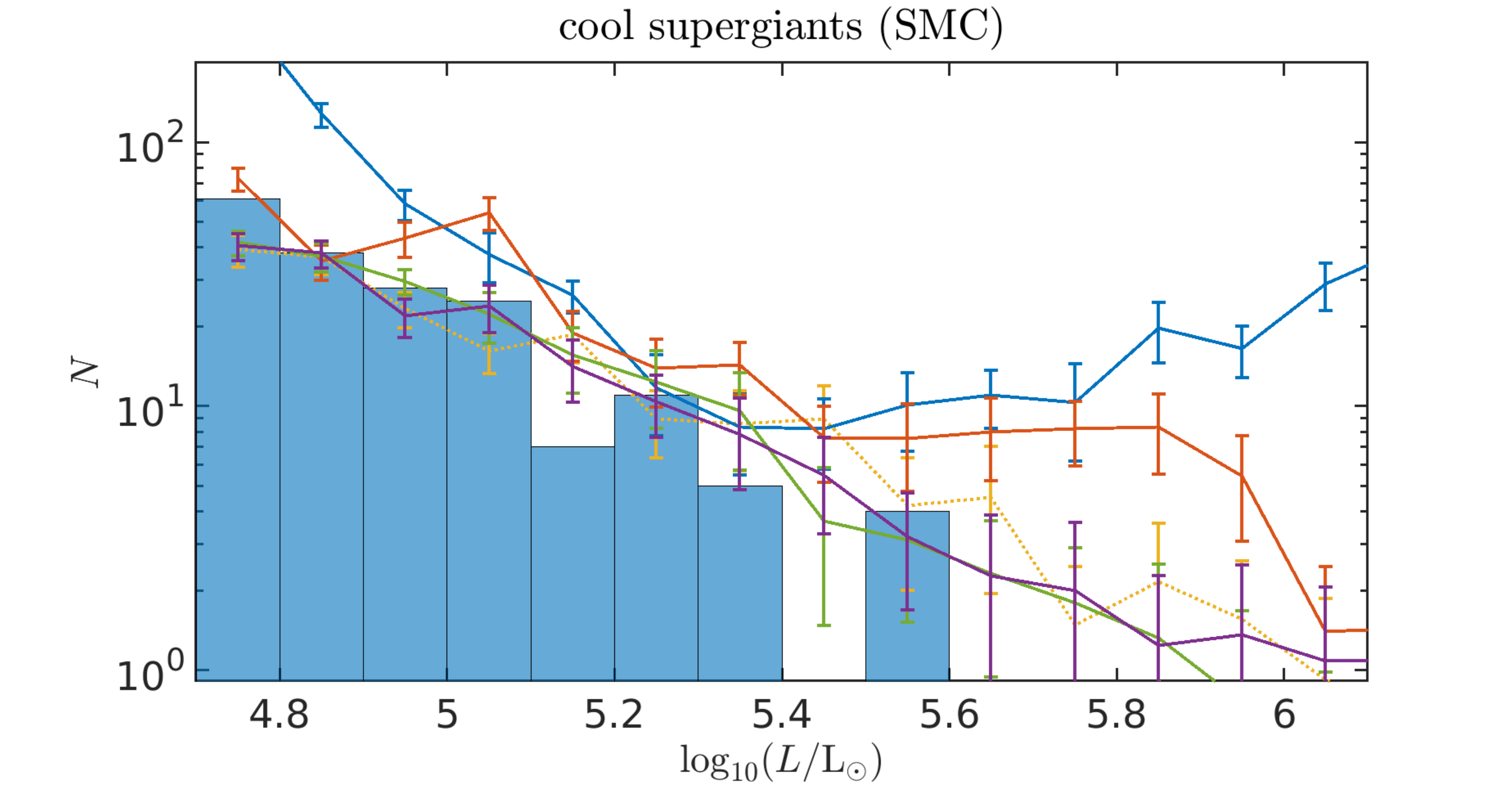} &
\includegraphics[width=0.465\textwidth]{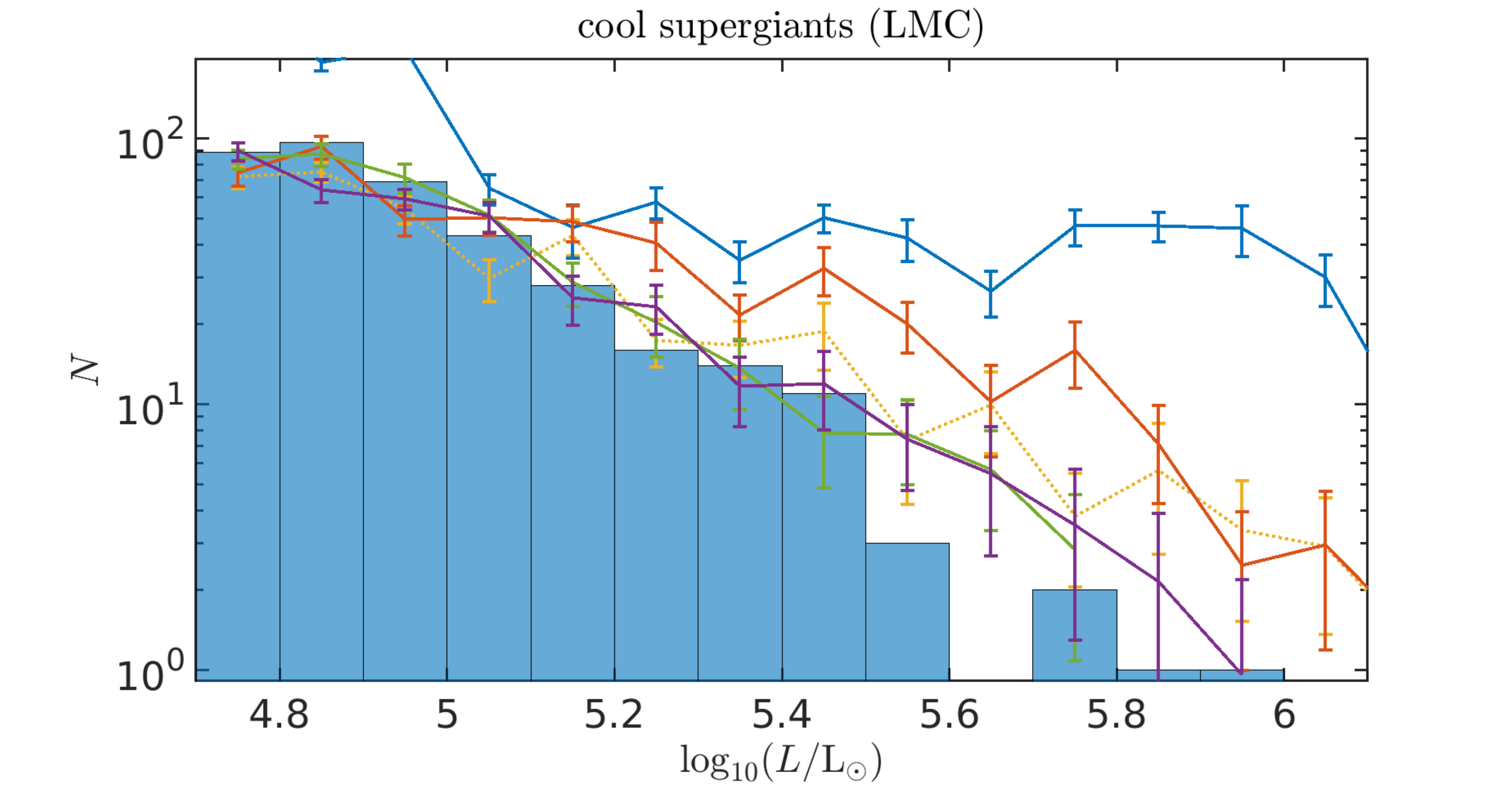} \\ 
\end{tabular}
\caption{Number of supergiant stars as function of luminosity, for a sample of five different sets of mixing parameters compared to the observed distributions for the SMC (left) and the LMC (right). The top panels show the number of RSGs ($T_\mathrm{eff} \le 4\, 800 \, \mathrm{K} $), the middle panels show RSGs and YSGs ($T_\mathrm{eff} < 7\, 500 \, \mathrm{K} $), and the bottom panels show cool ($T_\mathrm{eff} \le 12\, 500 \, \mathrm{K} $) supergiant stars.}
\label{fig:popsyn1vs100lum}
\end{figure*}

In Figure \ref{fig:popsyn1vs100lum} we show the simulated luminosity distribution for a few sets of mixing parameters compared to the observed distribution. For the RSGs, the simulated distribution is similar to the observed distribution for most mixing assumptions, with a small excess in the simulations for some cases, such as the combination of $\alpha_\mathrm{ov}=0.335$ and $\alpha_\mathrm{sc}=1$. The excess in the simulations increases when including also the YSGs, especially as there is a non-negligible number of such stars in the simulated populations, while the number of observed luminous YSGs is small. When including also cool BSGs, i.e. all stars with $T_\mathrm{eff} \le 12\, 500\, \mathrm{K}$, the excess of over-luminous stars in the simulated populations becomes even larger. An excess in the lower luminosities for BSGs is acceptable as we do not expect the sample to be complete for these temperatures, but higher luminosity stars ($\log_{10}\left(L/ \mathrm{L}_\odot\right) \ge 5.4$) would definitely be observed, and therefore the simulations cannot be taken to properly account for the stellar population.

\begin{figure}
\centering
\begin{tabular}{c}
\includegraphics[width=0.465\textwidth]{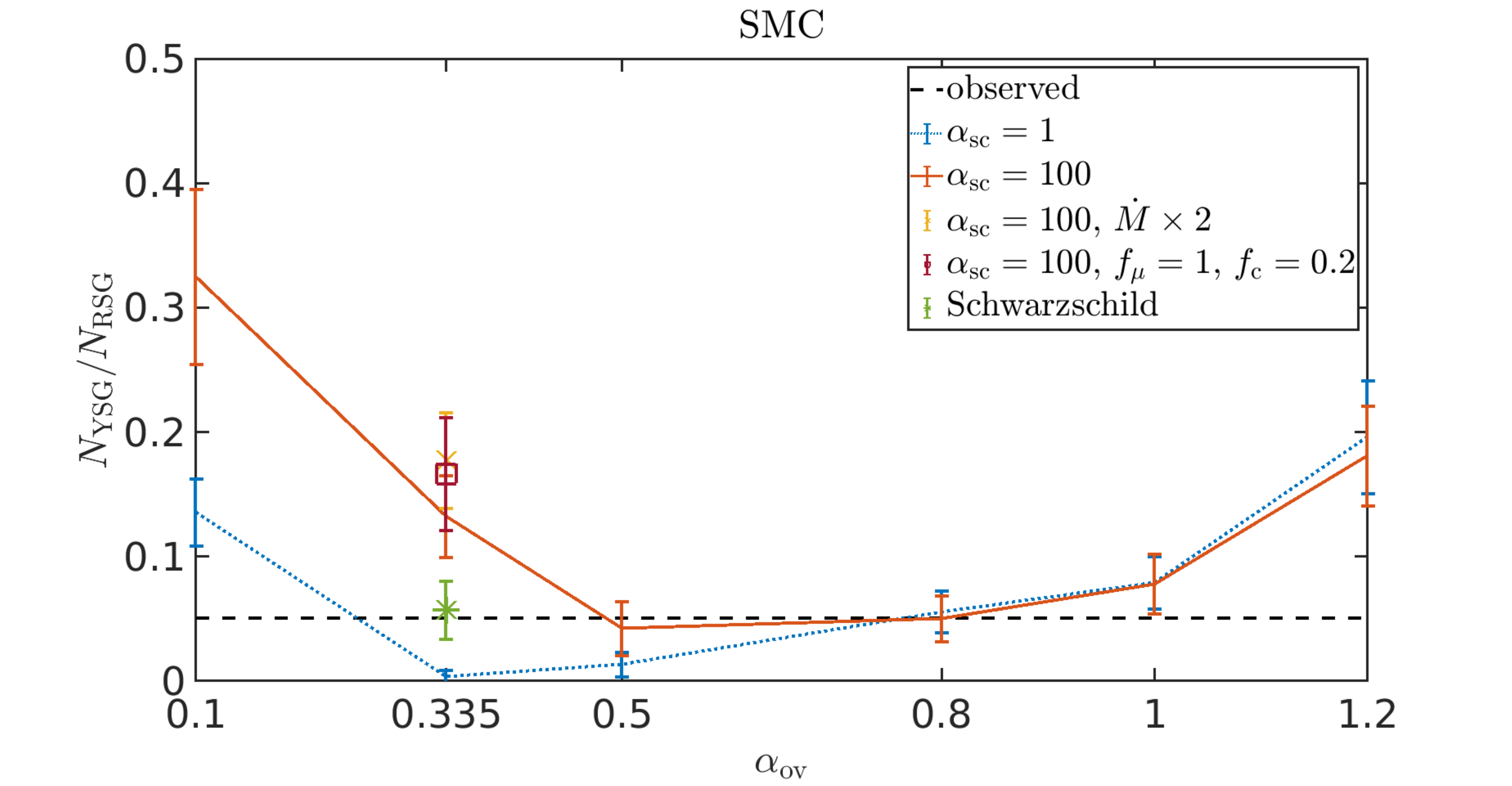} \\ 
\includegraphics[width=0.465\textwidth]{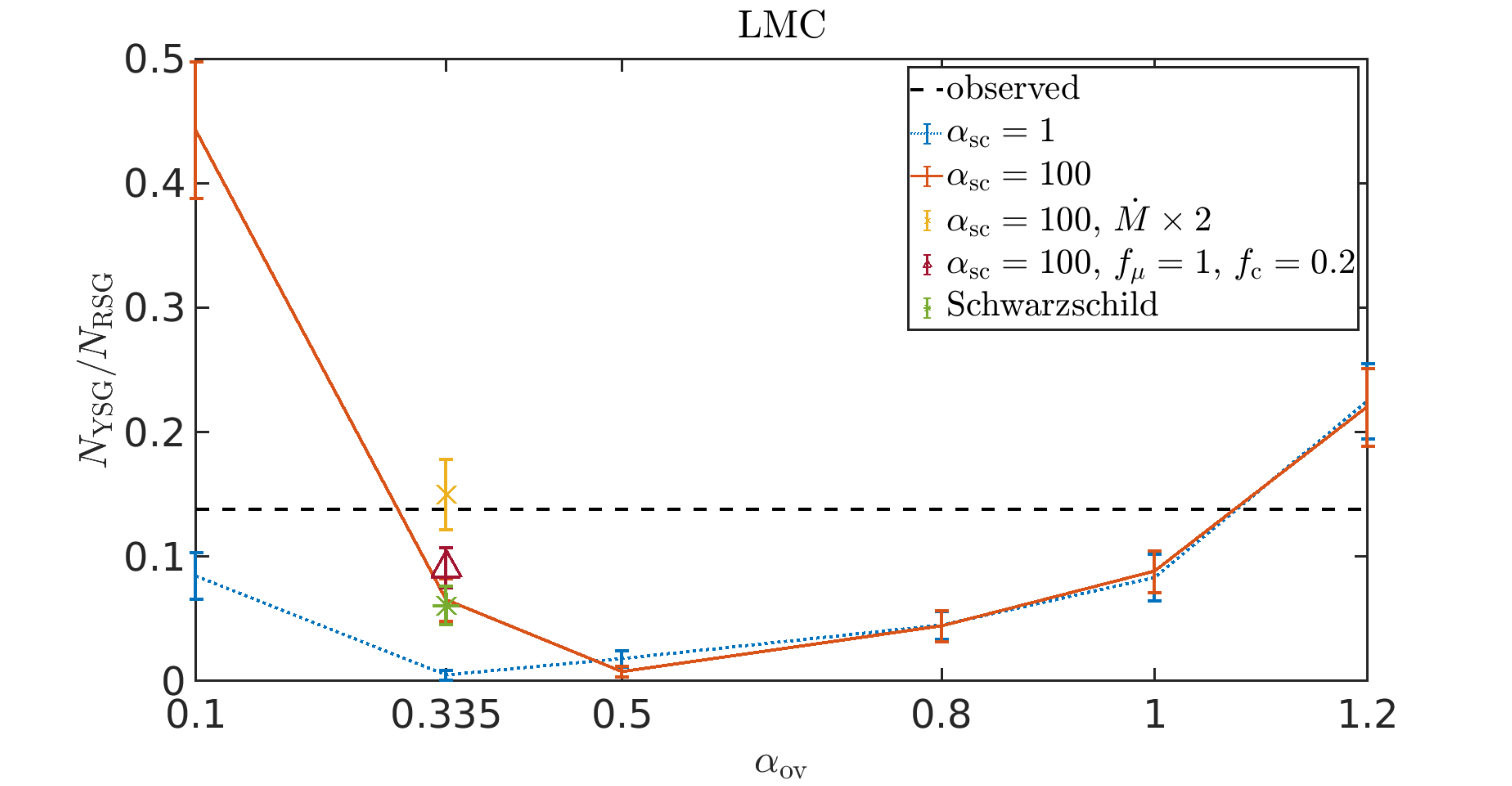} \\ 
\end{tabular}
\caption{Number of YSG stars relative to the number of RSG stars in synthetic populations as function of the overshooting extent, for two values of the semiconvective mixing efficiency, for the SMC (top) and the LMC (bottom).}
\label{fig:popsyn1vs100ratio}
\end{figure}

In Figure \ref{fig:popsyn1vs100ratio} we show the ratio between the number of RSGs with $4.7 \le \log_{10}\left(L / \mathrm{L}_\odot\right) < 5.4$ to the number of YSGs in the same luminosity range. This ratio is a good test for mixing in evolved massive stars. \cite{Schootemeijeretal2019} discuss a similar ratio, but between the numbers of RSGs and BSGs. We note also that the expansion and colour of supergiants depends on the treatment of convection, and employing MLT++ results in very luminous stars being hotter \citep{Klenckietal2020}. In the present study we focus on the most luminous stars, whose numbers are small compared to those with lower luminosities which mostly affect the YSG/RSG (or BSG/RSG) ratio. Supergiants of all colours need to be taken into account, to separate the issue of the number ratio of different colours from the issue of the HD limit.

\section{Sources of uncertainty and their impact}
\label{sec:assumptions}

\subsection{Mass loss}
\label{subsec:massloss}

Alongside mixing, continuous or eruptive mass loss is a key parameter that is responsible for stripping a star of its H-rich envelope. The larger the mass loss, the less likely the star is to cross the HD limit. Motivated by the apparent $Z$-independence of the HD limit, we chose to focus on mixing here.   However, to shed more light on the interplay between mass loss and mixing in this context, we also provide a set of models with mass-loss rates that are boosted by a factor of two. Recent theoretical and empirical determinations of mass-loss rates during the OB and the RSG phases \cite[e.g.,][]{ramachandran_testing_2019, Bjorklund2020, Beasor2020} rather suggest that standard prescriptions such as those used here already lead to an overestimation of the mass loss. Nevertheless, considering the poor understanding of eruptive mass loss, we explore the impact of an increased mass loss. We do this for the case of $\alpha_\mathrm{ov} = 0.335$ and $\alpha_\mathrm{sc} = 100$. 

The impact of boosting the mass loss throughout the stellar evolution by a factor of two is shown in Figure \ref{fig:popsyn1vs100}. As could be anticipated, boosting the mass loss helps reduce the excess of CSGs above the HD limit. However, even with increased mass loss, the discrepancy remains significant. Moreover, while it leads to an appreciable improvement in the case of the LMC, the improvement is negligible in the SMC, which again opposes the apparent $Z$-independence of the HD limit. To conclude, while continuous mass loss will play a role in shaping the HD limit, there is little support that it alone can explain the observations. It is well possible that a $Z$-independent rapid phase of mass loss (e.g., in a CSG or LBV phase) needs to be invoked to avoid the discrepancy between observations and theory, but a consistent physical framework for implementing it is still missing. 

\subsection{Mixing prescriptions}
\label{subsec:mixingpres}

The main uncertainty in stellar evolution modelling on which we focus in this work is mixing in stellar interiors. Therefore, in addition to the efficiency of semiconvection and the extent of overshooting, the effects of a couple of other mixing assumptions were tested. While we mostly used the Ledoux criterion for defining the convective boundary, in one set of models (with $\alpha_\mathrm{ov}=0.335$) we used the Schwarzschild criterion instead. The stability criterion employed affects not only the boundary of the convective core, but also intermediate regions between the core boundary and the stellar photosphere and the position of models in the HRD \citep*{Georgyetal2014}. The analysis of our synthetic populations generated from stellar evolution tracks which employed the Schwarzschild criterion yields a somewhat reduced excess of over-luminous CSGs compared to the analysis with the Ledoux criterion and the same overshooting extent. Similarly to our results with boosted mass loss (Section \ref{subsec:massloss}), the excess remains non-negligible, especially when considering the entire relevant temperature range.

Up until now we have discussed mixing only for convective regions and near their boundaries. In models of rotating stars, considerable mixing occurs also in radiative regions, owing to various instabilities \cite[e.g.,][]{Hegeretal2000}. In one set of models we changed the rotational mixing parameters in radiative regions to $f_c = 0.2$ and $f_\mu=1$, as described in Section \ref{subsec:mixrot}. Our analysis of the synthetic populations generated from tracks with these alternative mixing parameters yields almost no change in the CSG excess (Fig. \ref{fig:popsyn1vs100}). This might be a result of the possible degeneracy of these parameters \citep{ChieffiLimongi2013}.

\begin{figure}
\centering
\begin{tabular}{c}
\includegraphics[width=0.465\textwidth]{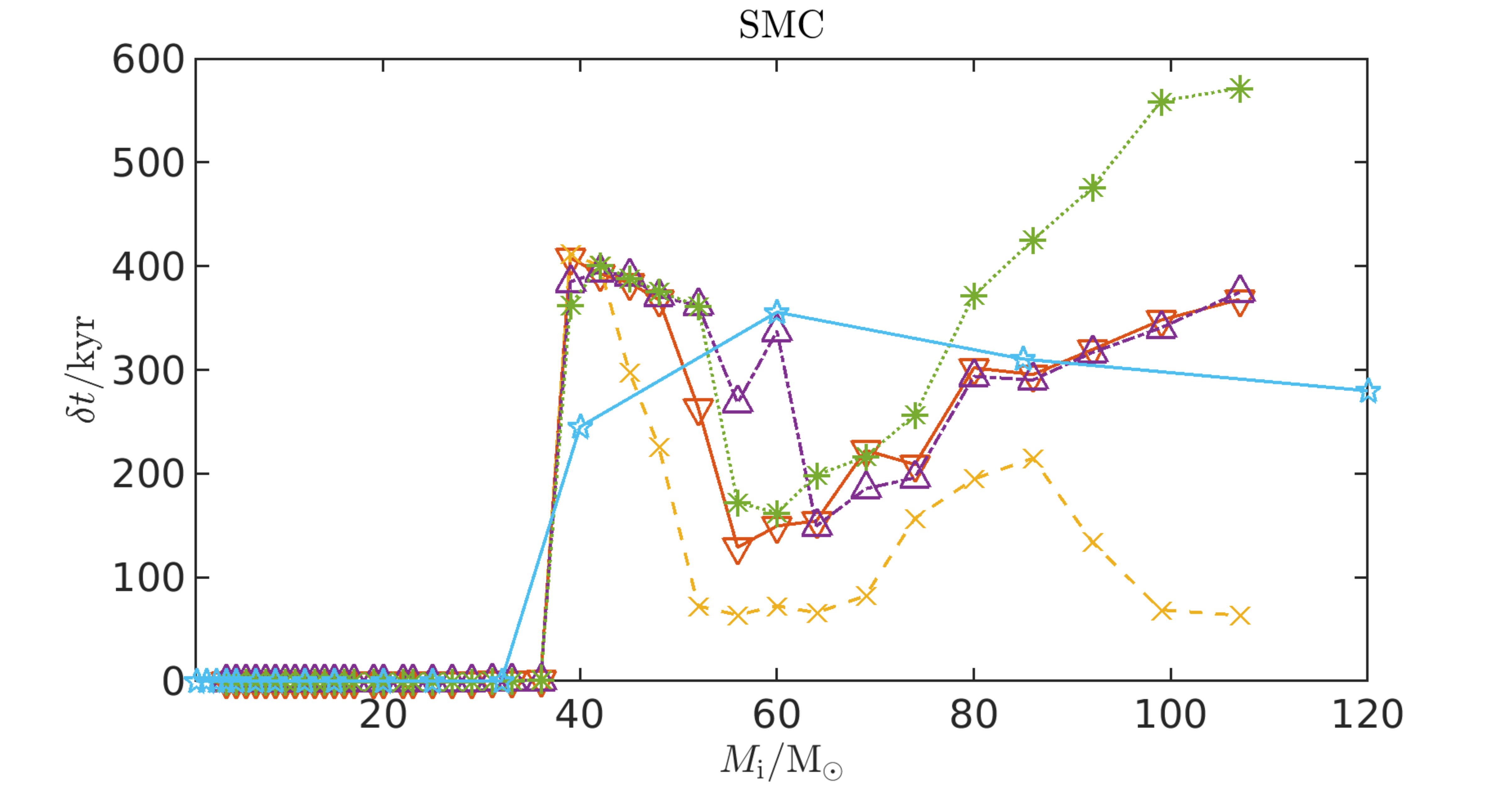} \\ 
\includegraphics[width=0.465\textwidth]{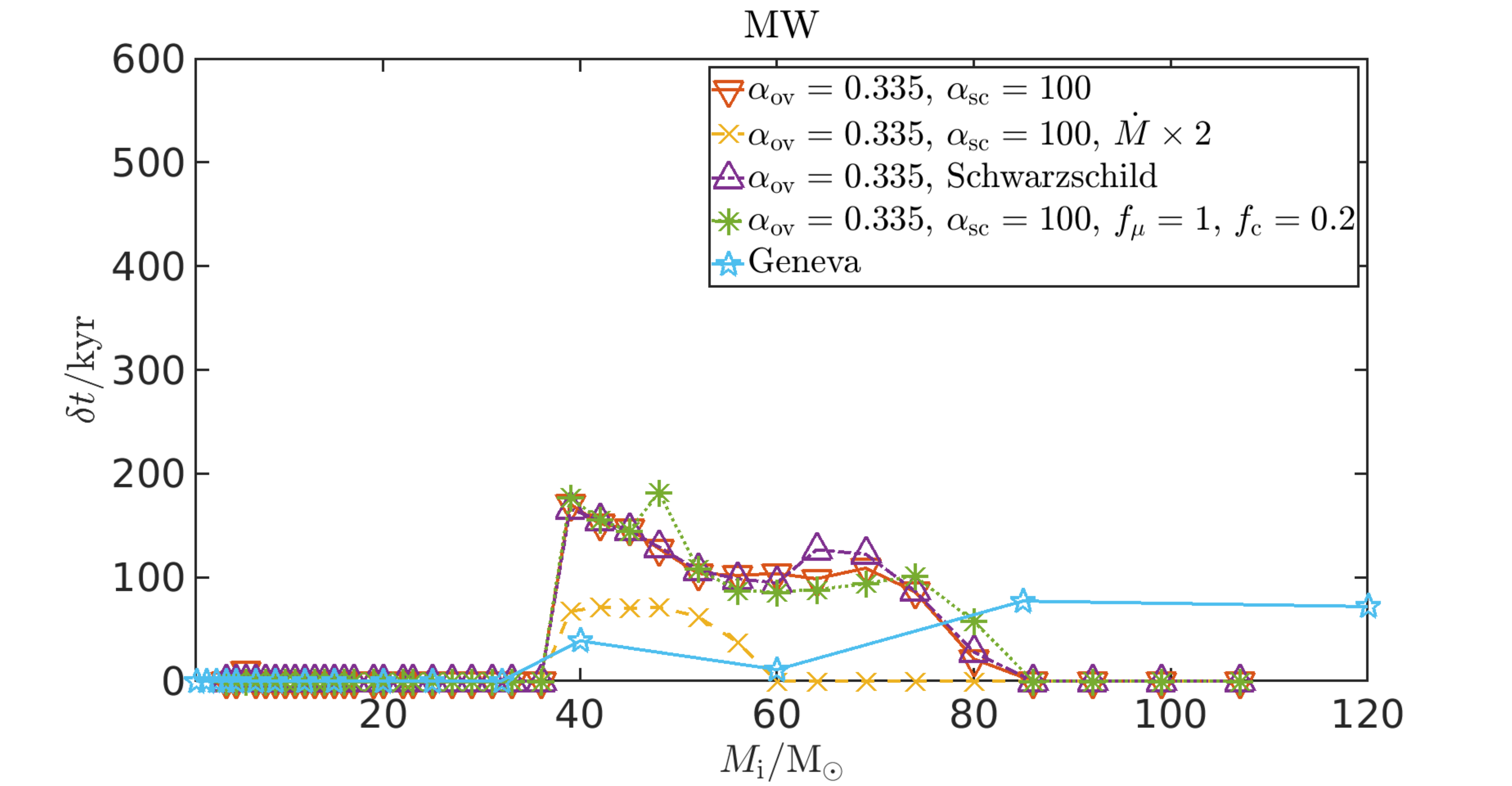} \\ 
\end{tabular}
\caption{Time spent beyond the HD limit as function of ZAMS mass for SMC (top) and MW (bottom) composition, for stellar evolution tracks as described in the inset.}
\label{fig:dtHD2}
\end{figure}

As an additional test of the sensitivity to the implementation of rotational mixing, we also compare our results to the Geneva tracks with MW and SMC compositions \citep{Ekstrometal2012, Georgyetal2013}. In Figure \ref{fig:dtHD2} we show the time spent beyond the HD limit for several of our modelling assumptions and for the Geneva models, for which rotational mixing is treated as an advective-diffusive process, compared to the diffusion approximation adopted by \texttt{MESA}. The Geneva models spend time periods beyond the HD limit rather similar to our tracks computed with \texttt{MESA}. While a quantitative comparison using synthetic populations will give a more definitive answer, we surmise that the computed excess of CSGs will be similar with the Geneva tracks.

\subsection{Multiplicity}
\label{subsec:multiplicity}

In principle, the presence of a close companion (orbital period $P \lesssim 3\,\mathrm{yr}$) would inhibit the expansion of its companion. Because a large fraction of massive stars interact with a companion during their evolution \citep{Sana2012,MoeDiStefano2017}, the effects of binaries on the simulated populations of CSGs and the HD limit should be taken into account. Given the many uncertain parameters that describe the properties and evolution of binaries, we decided to not account for multiplicity here. One may naively expect that binary interactions may be highly relevant for explaining the HD limit, since they can prevent stars from becoming CSGs. However, because the total number of RSGs and YSGs is used for the normalisation of our synthetic populations, and because it itself will be affected by binary interactions, the effect of stellar multiplicity is not trivial. If the binary separation shrinks as the binary mass increases, then one may expect binary interactions to help resolve the discrepancy. However, solid evidence for this at the upper-mass end is currently lacking. Addressing this question in future studies should be helpful to quantitatively constrain the impact of multiplicity on the HD limit.

\subsection{Star-formation history}
\label{subsec:SFH}

There are strong indications that the recent star-formation history in the Magellanic Clouds, especially in the SMC, was not continuous but is rather characterised by various peaks over the past $100\,\mathrm{Myr}$ \citep{Indu2011, ramachandran_testing_2019, Fulmer2020}. As such, this can have an important impact on the synthetic populations obtained, and hence on the final conclusions. However, it is unlikely that star-formation history alone can explain the discrepancy in the SMC and the LMC simultaneously. Moreover, the presence of WR stars in both galaxies that span a substantial luminosity range \citep{Hainich2014, Hainich2015, Shenar2016, Shenar2019} suggests that a similar distribution of CSGs across the luminosity range could be anticipated. Given the many uncertainties, repeating this investigation with detailed star-formation histories is beyond the scope of our work, but should be explored in future studies.

\section{Summary and discussion}
\label{sec:discussion}

We evolve numerous grids of stellar evolution tracks, for MW, LMC and SMC compositions, with a variety of mixing prescriptions. We find that enhanced mixing diverts stellar evolution tracks from the ``forbidden region'' defined by the HD limit (Figs. \ref{fig:HR1}-\ref{fig:HR2}). Based upon these grids of stellar models we construct synthetic populations with initial compositions appropriate for the LMC and SMC, and make quantitative comparisons with the observed populations of cool ($T_\mathrm{eff} \le 12.5\,\mathrm{kK}$) supergiant stars in these galaxies (Section \ref{sec:popsynvsobs}). We find that enhanced mixing reduces the excess of over-luminous CSGs in our simulated stellar populations. While for RSGs there does not seem to be a severe problem, the tension between observations and simulations increases with increasing the upper temperature cutoff. We can therefore consider the existence of a ``Cool Supergiant  Problem'' as the apparent mismatch between observed supergiants and the results of stellar evolution models.
 
When considering only RSGs, we find that the dependence of the excess on the overshooting extent is non-monotonic for efficient semiconvective mixing ($\alpha_\mathrm{sc}=100$). Recent studies by \cite{Schootemeijeretal2019} and \cite{HigginsVink2020} advocate for highly efficient mixing in regions of semiconvection to account for the properties of supergiants in the LMC and SMC. \cite{Schootemeijeretal2019} also claim that convective overshooting can be constrained by the properties of the populations, such as the ratio between RSGs and BSGs. We differ from \cite{Schootemeijeretal2019} in two regards: ($i$) We consider much higher values of $\alpha_\mathrm{ov}$, beyond an apparent extremum around $\alpha_\mathrm{ov} \approx 0.5$; ($ii$) \cite{Schootemeijeretal2019} focus on the bulk of the supergiant population, while we investigate the stars with the highest luminosities, which are a small fraction of the overall population.

\cite{HigginsVink2020} suggest that the HD limit can be explained with decreased overshoot mixing, $\alpha_\mathrm{ov}=0.1$, though they consider only the red part of the evolution. We show that for red supergiants taking $\alpha_\mathrm{ov}=0.1$ together with $\alpha_\mathrm{sc}=100$ indeed gives a reasonable account of the supergiant populations in the LMC and SMC (Figs. \ref{fig:popsyn1vs100}-\ref{fig:popsyn1vs100lum}). This is explained by the significant effect of efficient semiconvection in increasing the effective surface temperatures of the stellar models. Thus, for efficient semiconvection, core He-burning stars tend to appear as YSGs or BSGs instead of RSGs. However, this does not solve the supergiant excess beyond the HD limit, but merely ``sweeps it under the carpet''. When we  consider the entire temperature range to which the horizontal HD limit applies ($T_{\rm eff} \lesssim 12.5\,\mathrm{kK}$) we find that there is a large excess of stellar models in our simulated populations. As \cite{HigginsVink2020} did not consider the same temperature range as us, there is no discrepancy between the results. It is important to realise that the HD limit is not limited only to RSGs, and hence all relevant temperatures need to be included.

It is also evident from our results that the extent of overshooting affects the formation of WR stars. Enhanced mixing reduces the mass threshold for a star to remove its envelope through winds, and therefore affects the necessity of binary interactions in forming WR stars \citep{Shenaretal2020}, as well as decreasing the single-star evolution WR luminosity threshold. For example, Figs. \ref{fig:HR1}-\ref{fig:HR2} show that increasing $\alpha_\mathrm{ov}$ from $0.335$ to $1.2$ reduces the lowest luminosity of a WR star formed through single-star evolution from $\log_{10} \left( L / \mathrm{L}_\odot\right) \approx 6$ to $\approx 5.5$ for the LMC, and from $\approx 5.5$ to $\approx 5.25$ for the MW.

Our results also have implications for energetic transient astrophysical phenomena. Firstly, the initial mass threshold for core-collapse supernovae is lower for enhanced mixing. Secondly, the position in the HRD where stellar models end their lives is substantially hotter for enhanced mixing. This has implications for the so-called ``Red Supergiant Problem'', which is the apparent discrepancy between the most luminous progenitor of a Type IIP supernova and the most luminous known RSGs \citep{Smarttetal2009, Horiuchi2014, DaviesBeasor2018, DaviesBeasor2020, Kochanek2020}. Our models with lower overshooting values exhibit such a problem, though for enhanced mixing the more massive stars do not die as RSGs, in accordance with observations. 

The formation of black hole binaries (which are progenitors of gravitational wave events) depends on the expansion of massive stars and their interaction with a companion. \cite{Klenckietal2020} discuss the role of metallicity in stellar expansion and interaction, as with lower metallicity stars are generally more compact because of the lower gas opacity. Smaller stellar radii imply that a smaller fraction of stars will experience significant binary interactions, such as common envelope evolution, that lead to short-period binary black holes which will merge quickly enough to produce observable gravitational wave events. Our models with enhanced mixing expand to smaller radii in general, therefore affecting the evolution towards merging black holes. Since a large fraction of massive stars interact with a companion during their evolution

\cite{Klenckietal2020} also discuss the role of MLT++ in relation to the HD limit. According to \cite{Klenckietal2020}, the more limited expansion of massive stellar models employing MLT++ is favorable in terms of explaining the HD limit, and it might be more accurate for rather massive ($M \ga 50 \,\mathrm{M}_\odot$) stars. We note that even though we use the favorable MLT++ prescription we still find an excess of over-luminous stars in the temperature range relevant for the HD limit.

It is important to stress that our study focuses on the upper-mass end.  For example, applying high $\alpha_{\rm ov}$ values as invoked here can suppress blue loops \citep*{Stothers1991,Walmswell2015} on the HRD (Fig. \ref{fig:HR1}) and prevent lower-mass stars from becoming Cepheids \citep{Anderson2014,Anderson2016}, in contrast to observations. However, studies of intermediate-mass stars suggest a mass-dependent convective overshoot extent \citep{ClaretTorres2016,ClaretTorres2017,ClaretTorres2018,ClaretTorres2019}. These studies suggest an increase up to $\alpha_\mathrm{ov} \approx 0.2$ at $M\approx 2\,\mathrm{M}_\odot$, plateauing afterwards. A similar trend is explained by \cite{JTC}. Overshooting is less well-constrained for the higher masses which are relevant for our study and for the existence of the HD limit. Models of massive stars make various calibrations for the overshooting extent, with results ranging from $\alpha_\mathrm{ov}=0.1$ calibrated for the mass range of $1.3$-$9\,\mathrm{M}_\odot$ \citep{Ekstrometal2012} to $\alpha_\mathrm{ov}=0.5$ for higher masses around $M \ga 30\,\mathrm{M}_\odot$ \citep{HigginsVink2019}. So while very high overshoot values cannot be applied to the lowest masses in our grids ($M=4\,\mathrm{M}_\odot$), higher mass stars might exhibit behaviour appropriate for enhanced mixing, with few real constraints on the extent.

In conclusion, we propose that internal mixing in massive stars might play an important part in explaining the empiric HD limit. Enhanced mixing prevents the redward evolution of stellar models towards or beyond the HD limit. We do not suggest to adopt a higher overshooting parameter, but rather that our results hint at a deficiency in the modelling of mixing in stellar interiors \citep[e.g.,][]{Aerts2019, Schootemeijeretal2019}. The extent of core overshooting might be highly mass-dependent, or rotational mixing is more efficient for slowly rotating stars. Moreover, it is well possible that the final explanation relies on a multitude of mechanisms (e.g., mixing, mass loss, multiplicity, star formation history; see Sect.\,\ref{sec:assumptions}). We therefore encourage future investigations of this problem that address these various mechanisms. For now, internal mixing in massive stars remains an unresolved issue in stellar modelling with broad implications, and the origin of the Humphreys-Davidson limit remains uncertain.

\section*{Acknowledgments}
We acknowledge the constructive report by our referee Dr.\ Cyril Georgy.
This research has made use of the SIMBAD database and  cross-match service operated at CDS, Strasbourg, France. The research has received funding from the European Research Council (ERC) under the European Union's Horizon 2020 research and innovation programme with grant number 772225: MULTIPLES.
AG and TS would like to thank Danny Lennon and Ben Davies for helpful input and exchanges.
The Flatiron Institute is supported by the Simons Foundation.
IA is a CIFAR Azrieli Global Scholar in the Gravity and the Extreme Universe Program and acknowledges support from that program, from the European Research Council (ERC) under the European Union’s Horizon 2020 research and innovation program (grant agreement number 852097), from the Israel Science Foundation (grant numbers 2108/18 and 2752/19), from the United States - Israel Binational Science Foundation (BSF), and from the Israeli Council for Higher Education Alon Fellowship.

\section*{Data Availability Statement}
The data underlying this article will be shared on reasonable request to the corresponding author.

\bibliographystyle{mnras}
\input{CSGprob.bbl}

\label{lastpage}

\end{document}